\documentclass[%
 reprint,
 amsmath,amssymb,
 aps,
 prx,
floatfix,
]{revtex4-1}

\usepackage{xcolor}
\usepackage{graphicx}% Include figure files
\usepackage{dcolumn}% Align table columns on decimal point
\usepackage{bm}% bold math
\usepackage{hyperref}% add hypertext capabilities
\usepackage[mathlines]{lineno}% Enable numbering of text and display math
\begin{document}

\preprint{APS/123-QED}

\title{Universal reshaping of arrested colloidal gels via active doping}% Force line breaks with \\
%\thanks{A footnote to the article title}%

\author{S. A. Mallory}
\email{smallory@caltech.edu}
\affiliation{%
 Division of Chemistry and Chemical Engineering, California Institute of Technology, Pasadena, CA 91125
}%

%\collaboration{MUSO Collaboration}%\noaffiliation

\author{M. L. Bowers}
\author{A. Cacciuto}
\email{ac2822@columbia.edu}
\affiliation{Department of Chemistry, Columbia University, New York, New York 10027}%

%\collaboration{CLEO Collaboration}%\noaffiliation

\date{\today}% It is always \today, today,
             %  but any date may be explicitly specified

\begin{abstract}
Colloids that interact via a short-range attraction serve as the primary building blocks for a broad range of self-assembled materials. 
However, one of the well-known drawbacks to this strategy is that these building blocks rapidly and readily condense into a metastable colloidal gel.
Using computer simulations, we illustrate how the addition of a small fraction of purely repulsive self-propelled colloids, a technique referred to as active doping, can prevent the formation of this metastable gel state and drive the system toward its thermodynamically favored crystalline target structure.
The simplicity and robust nature of this strategy offers a systematic and generic pathway to improving the self-assembly of a large number of complex colloidal structures.
We discuss in detail the process by which this feat is accomplished and provide quantitative metrics for exploiting it to modulate self-assembly.
We provide evidence for the generic nature of this approach by demonstrating that it remains robust under a number of different anisotropic short-ranged pair interactions in both two and three dimensions.
In addition, we report on a novel microphase in mixtures of passive and active colloids. 
For a broad range of self-propelling velocities, it is possible to stabilize a suspension of fairly monodisperse finite-size crystallites. 
Surprisingly, this microphase is also insensitive to the underlying pair interaction between building blocks. 
The active stabilization of these moderately-sized monodisperse clusters is quite remarkable and should be of great utility in the design of hierarchical self-assembly strategies. 
This work further bolsters the notion that active forces can play a pivotal role in directing colloidal self-assembly.
\end{abstract}

%\keywords{Suggested keywords}%Use showkeys class option if keyword
                              %display desired
\maketitle

%\tableofcontents

\section{Introduction}

The directed self-assembly of a well-defined microscopic structure is a complex process that is difficult to control with any degree of precision.
This has prompted the ongoing challenge of designing colloidal building blocks which can spontaneously and deliberately organize into functional microstructures with desired material properties \cite{Glotzer2004a,  glotzer_assembly_2004, Sacanna2013a, Li2011, Sacanna2011, Zhang2015, Jankowski2012}. 
Many of the current design strategies rely on colloidal building blocks that interact via a short-ranged attraction where the range of interaction does not exceed the colloids own diameter.
Inspired by the precise self-assembly of protein complexes, virial capsids, and other molecular systems, it is becoming increasingly possible to drive the monodisperse self-assembly of specific colloidal structures by either altering the shape of the colloids or by introducing highly selective interactions.
Prominent examples of synthetic building blocks include: Janus and patchy particles \cite{Zhang2017,Ravaine2017,Morphew2018,Li2020}, lock and key colloids \cite{Ashton2013,Sacanna2010a,Wang2014}, and DNA decorated colloids \cite{Wang2012a,Biancaniello2005,Nykypanchuk2008,Valignat2005,Rogers2016}.   
This approach has been particularly fruitful, and a variety of colloidal materials with promising mechanical and optical properties have been synthesised in this manner over the past decade. 

As part of this effort, it has been well-documented that high yield self-assembly only occurs for very specific particle shapes and interaction strengths, making the search for the optimal self-assembly conditions rather cumbersome \cite{whitelam_role_2009, whitelam_statistical_2015, noauthor_exploiting_2010}.
For most colloidal systems that interact via sufficiently short-ranged pair interactions, there exist an ensemble of competing metastable or kinetically arrested structures (i.e. gels) that directly compete with the formation of the desired target structure \cite{Lin1989,Sciortino2004,Charbonneau2007,Lu2008,fernandeztoledano_colloidal_2009,Sanchez2014}. 
Unless the self-assembly pathway is fully understood and system conditions are highly optimized, the time required to form the target structure can become prohibitively long, if accessible at all. 

Through recent advances in colloidal synthesis, a new class of colloid has been developed with the unique capability of exploiting local energy or chemical gradients to propel themselves at speeds of tens of microns per second \cite{Yan2016,Wang2015,Gao2014,Brown2014,Palacci2013a,wang_small_2013,dey_chemically_2017}. 
These synthetic particles can be thought of as the colloidal analog of swimming bacteria. 
Crucially, a major benefit of these synthetic variants over their biological counterpart is the ability to systematically tailor interparticle interactions and dynamically modulate both the mechanism and speed of self-propulsion. 
The inherent non-equilibrium nature of these active colloids and their rich collective behavior have inspired a large body of work in the field of non-equilibrium statistical physics \cite{Ramaswamy2010,Bechinger2016,DiLeonardo2016,Patteson2016,zottl_emergent_2016,Cates2015,Dauchot2019,Speck2020,MariniBettoloMarconi2015}. 
From a materials engineering perspective, an active particle's ability to autonomously navigate complex microfluidic environments conjures up a host of appealing applications.
Due to their unique self-driven nature, and ability to sense their environment at a scale comparable to their size, active particles are potentially the ideal tool to manipulate and transport matter at the microscale.
It has already been shown by multiple research groups that active colloids are a powerful medium for mediating the effective interactions between suspended microstructures \cite{Mallory2017a,Das2019,Shan2019,Prymidis2015,solovev_self-propelled_2012,Wang2019a,Angelani2019, Grunwald2016,maggi_self-assembly_2016, wang2019interactions}. 

In the context of active colloidal self-assembly, a series of recent studies have shown that a judicious use of self-propulsion can be beneficial when self-assembling both finite-size and macroscopic crystalline structures from single colloidal units \cite{Mallory2019,Wensink2014,Aldana2020,Mallory2017,Mallory2016,Hess2006,wykes_dynamic_2016,Du2019,ni2013pushing}.
The central theme throughout these studies is that by self-propelling each individual building blocks it is possible to drive the system away from some metastable, kinetically frustrated state into the desired final configuration.
The success of such a strategy hinges on the ability of controlling simultaneously both the strength and anisotropy of interparticle interactions as well as the strength and direction of the active forces for all self-assembling building blocks.
There has been some   important experimental work  in this direction \cite{Wang2019,Gao2013a,Yan2016}. 
However, it is a difficult task to synthesize colloids with full control over both the nature of the pair interaction and active forces. 

In this work, we leverage an alternative technique which circumvents this issue, and demonstrates that excellent self-assembly can be achieved by simply adding a relatively small number of purely-repulsive active particles to a solution of complex colloidal building blocks.
This strategy of active doping has been explored  both experimentally and theoretically by several research groups \cite{PhysRevLett.120.268004, C5SM00827A, lozano2019active,Omar2019,C7SM01730H, C4SM01015A, C9SM01496A, PhysRevLett.119.058001,C6SM00031B,Ramananarivo2019, C6SM00700G}. 
It has been shown to be an effective strategy for a range of microscopic task ranging from healing defects in colloidal crystals to modulating the structure of isotropic colloidal gels and glasses. 
Prior active doping studies have predominately focused on dense suspensions of passive colloidal building blocks that interact through isotropic pair interactions. 
One of the central findings in this study, is that active doping is insensitive to the details of the pair interaction between passive colloidal building blocks as long as it is sufficiently short-ranged. 
The significance of this cannot be overstated as these are the class of building block required to build complex colloidal structures that go beyond simple closed packed geometries. 

We apply the active doping strategy to three different colloidal self-assembly problems to illustrate that it provides a systematic and generic pathway to improving the self-assembly of a large number of complex colloidal structures.
The first is the formation of a hexagonal crystal using colloids with a short-ranged isotropic attraction as the building block.
The second example are triblock Janus colloids which are designed to self-assembly into a colloidal kagome lattice.
Even this difficult self-assembly problem of constructing an open-cell structure can be significantly enhanced using the protocol we detail. 
The final example is the formation of a two dimensional cubic crystal with square symmetry formed with patchy colloids equipped with four orthogonal attractive patches.
This simple strategy offers a higher level of control over system conditions in comparison to methods where each individual building blocks is active. 
We should also stress that our results can be easily tested experimentally as purely-repulsive active colloids are readily available experimentally. 
Lastly, we uncovered a previously unreported microphase in mixtures of passive and active colloids.  
For a range of self-propelling velocities, it is possible to stabilize a suspension of fairly monodisperse finite-size crystallites (See Fig. 1C).  
Interestingly, the stability of this phase also seems to be insensitive to the underlying pair interaction between passive building blocks.
We discuss the physical mechanism underlying the stabilization of this novel phase.

\section{Model}

Both active and passive colloids are modeled as spheres of diameter $\sigma$ that undergo Brownian dynamics at a constant temperature $T$ according to the coupled overdamped Langevin equations: 

\begin{equation}
\dot{\bm{r}}(t) = U \, \bm{n}(t) + \sqrt{2D}\,\bm{\xi}(t) + \gamma^{-1} \bm{F}_{ij}
\label{BDT}
\end{equation}
\begin{equation}
\dot{\bm{{n}}}(t)=  \sqrt{2D_R}\, \bm{\xi}_R(t) \times \bm{n}(t) + \gamma_R^{-1}\bm{T}_{ij}
\label{BDR}
\end{equation}

\noindent  where the translational diffusion of the colloid is given by the Stokes-Einstein relation $D=k_{\mathrm B}T\gamma^{-1}$ with $\gamma$ being the translational friction due to the fluid.
Unless stated otherwise, it can be assumed that all colloids are confined to move in the xy-plane (including the orientation vector of all active colloids). 
Each active colloid swims at a fixed speed $U$ along a predefined orientation unit vector $\bm{n}(t)$, while for all passive colloids $U=0$. 
For spherical colloids in a low Reynolds number environment it is reasonable to assume rotational diffusion is thermal in origin and given by $D_R=k_{\mathrm B}T\gamma_R^{-1}=3D\sigma^{-2}$ where the rotational friction is given by $\gamma_R$.
The random Brownian forces and torques acting on each colloid are modeled as Gaussian white-noise and are characterized by $\langle \bm{\xi}(t)\rangle = 0$ and $\langle \xi_\alpha(t) \xi_\beta(t^\prime)\rangle = \delta_{\alpha \beta}\delta(t-t^\prime)$ and $\langle \bm{\xi}_R(t)\rangle = 0$ and $\langle \xi_{R \alpha}(t) \xi_{R \beta}(t^\prime)\rangle = \delta_{\alpha \beta}\delta(t-t^\prime)$, respectively.
The interparticle forces and torques that arise from the chemical patterning of the colloid's surface are given by $\bm{F}_{ij}$ and  $\bm{T}_{ij}$, respectively. A more complete discussion of the different pair potential and their functional form can be found in the Supplemental Material. 

All simulations consist of a binary mixture of $N$ attractive passive colloids and $M$ purely repulsive active colloids confined to a periodic box of dimension $L$. 
We considered three different types of passive colloidal building blocks: 1) spherical particles with a steep short range attraction,  2) triblock Janus colloids with two attractive patches located at the poles such that they self-assembly into a kagome lattice, and 3) spherical particles equipped with four attractive patches oriented in such a way to form a square crystal in two dimensions.
As we are interested in the low-dosage regime where $M\ll N$, we consider ratios $\chi=M/N$  in the range  of 0.05-0.2.
The length and energy scale are chosen to be $\sigma$ and $k_{\mathrm B}T$, respectively, while $\tau=\sigma^2D^{-1}$ is the unit of time.
All simulations were run for $10^9-10^{10}$ timesteps with a timestep of $\Delta t=10^{-5}\tau$.
All results are reported in reduced Lennard-Jones units. As a useful reference, an active colloid in our simulations with swim speed $U=1$ corresponds to a swim speed of $\approx 1 \  \mu$m/sec for an active colloid with a diameter of $\approx 1 \  \mu$m.
All system snapshots were rendered with the OVITO visualization package \cite{Stukowski2010}.
The data that support the findings of this study are available from the corresponding author upon reasonable request.

\section{Results \& Discussion}

\begin{figure}[t!]
\centering
\includegraphics[width=.475\textwidth]{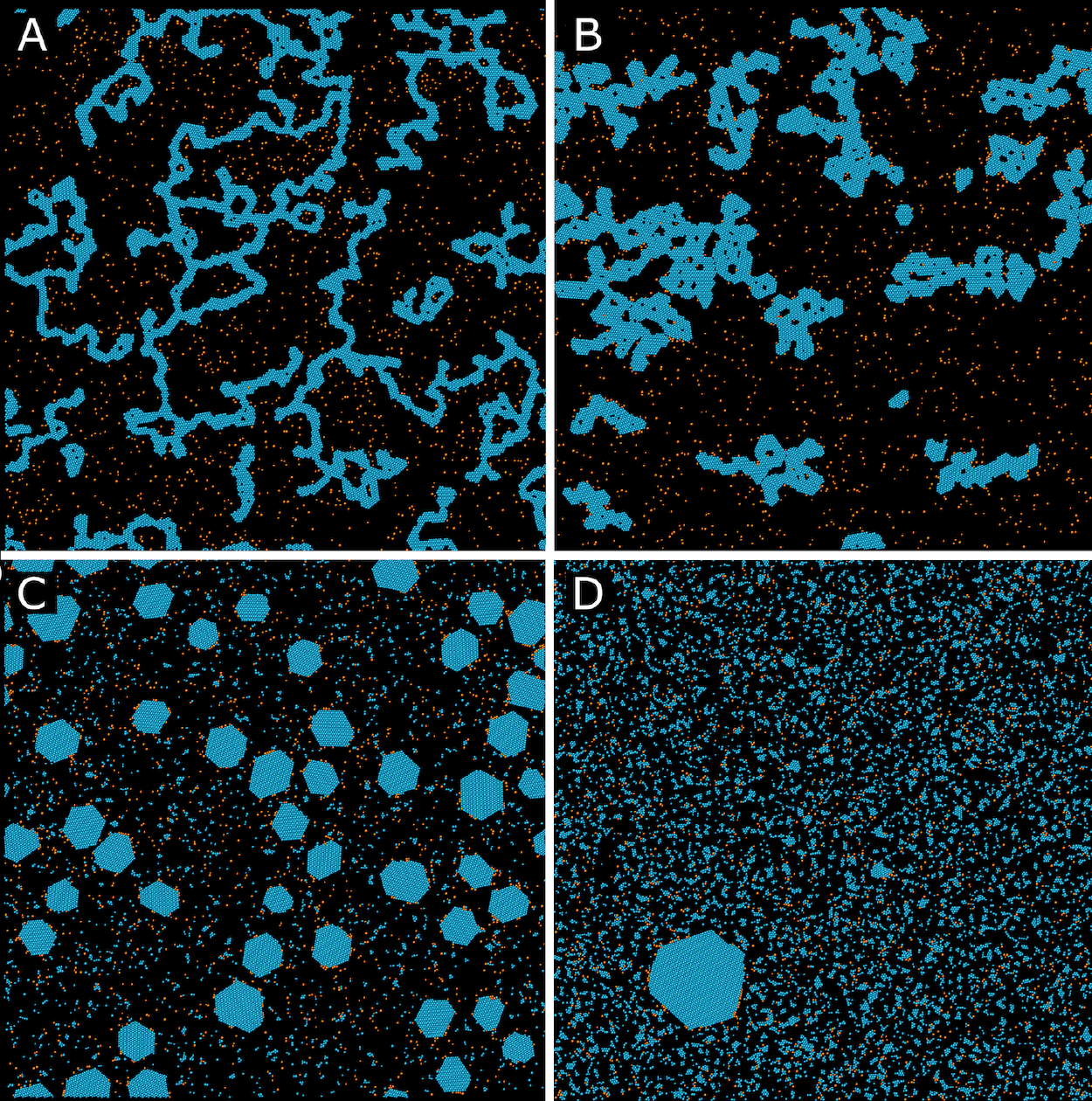}
\caption{Representative snapshots of the isotropic system with short-ranged attraction for different state points. The total volume fraction is fixed at $\phi=0.2$, and the ratio of purely repulsive active colloids $\chi=0.1$ for a total of 16384 particles. (A) Snapshot of the gel-like network characteristic of the equilibrium system $(U=0)$. (B) Collapsed gel-like network for low to moderate self-propelling velocities $U=20$  (C) Stable microphase of a fairly monodisperse suspension of crystallites for $U=100$, and (D) a single crystal coexisting with its own fluid for  $U=127$. The system completely melts for higher velocities. The passive particles are depicted with the color blue, while the active colloids are shown in orange. }
 \label{figure1}
\end{figure}

As previously noted, finding the thermodynamic parameters for which self-assembly is optimal can be quite cumbersome. 
The geometry and strength of the interactions need to be carefully engineered to ensure that once a cluster begins to form it can locally reorganize to heal any local defect that may get trapped in the growing structure.
Usually, the window in this parameter space where self-assembly is successful is rather narrow.
Even for a simple colloidal suspension with particles interacting via a generic short-range attraction, optimal self-assembly is obtained only when $\varepsilon/k_{\mathrm B}T \gtrsim 1 $.
In fact, when thermal forces dominate $(\varepsilon/k_{\mathrm B}T \ll 1)$, the system structurally remains in a gas phase where only small and short-lived clusters are observed.
In the opposite limit, when attractive forces dominate $(\varepsilon/k_{\mathrm B}T \gg 1)$, the system can easily become trapped in a malformed, metastable gel-like configuration.
This simple picture of self-assembly does not rely on microscopic system-specific details, provided the geometry of the interparticle interactions is compatible with the desired target structure, and thus applies broadly to any colloidal system with a short-range attractions.

To understand the overall effect a small number of purely-repulsive active particles has on the stability of a metastable gel, we first consider the case of a small fraction of purely-repulsive active colloids added to a suspension of passive colloids interacting via an isotropic short-ranged attraction.
This is implemented using a Mie potential (i.e. generalized Lennard-Jones) of the form

\begin{equation}
V(r_{ij})=4\varepsilon\left[ \left(\frac{\sigma}{r_{ij}}\right)^{36} - \left(\frac{\sigma}{r_{ij}}\right)^{18}  +\frac{1}{4}\right]
\label{potential}
\end{equation}

\noindent For passive colloids, the potential is cut off at a distance of $r_c=1.5\sigma$ and $\varepsilon = 6\, k_{\mathrm B}T$. 
This gives rise to a rather short-ranged attraction that favors the formation of kinetically frustrated gel-like structures. 
The active colloids that we add in solution are purely repulsive  and only interact via their excluded volume.
This purely repulsive interaction is realized by setting the cutoff of the  potential in Eq.~\ref{potential} to $r_c=2^{\frac{1}{18}}\sigma$ for all pair interactions involving active colloids. 

\begin{figure}[t!]
 \centering
    \includegraphics[width=.475\textwidth]{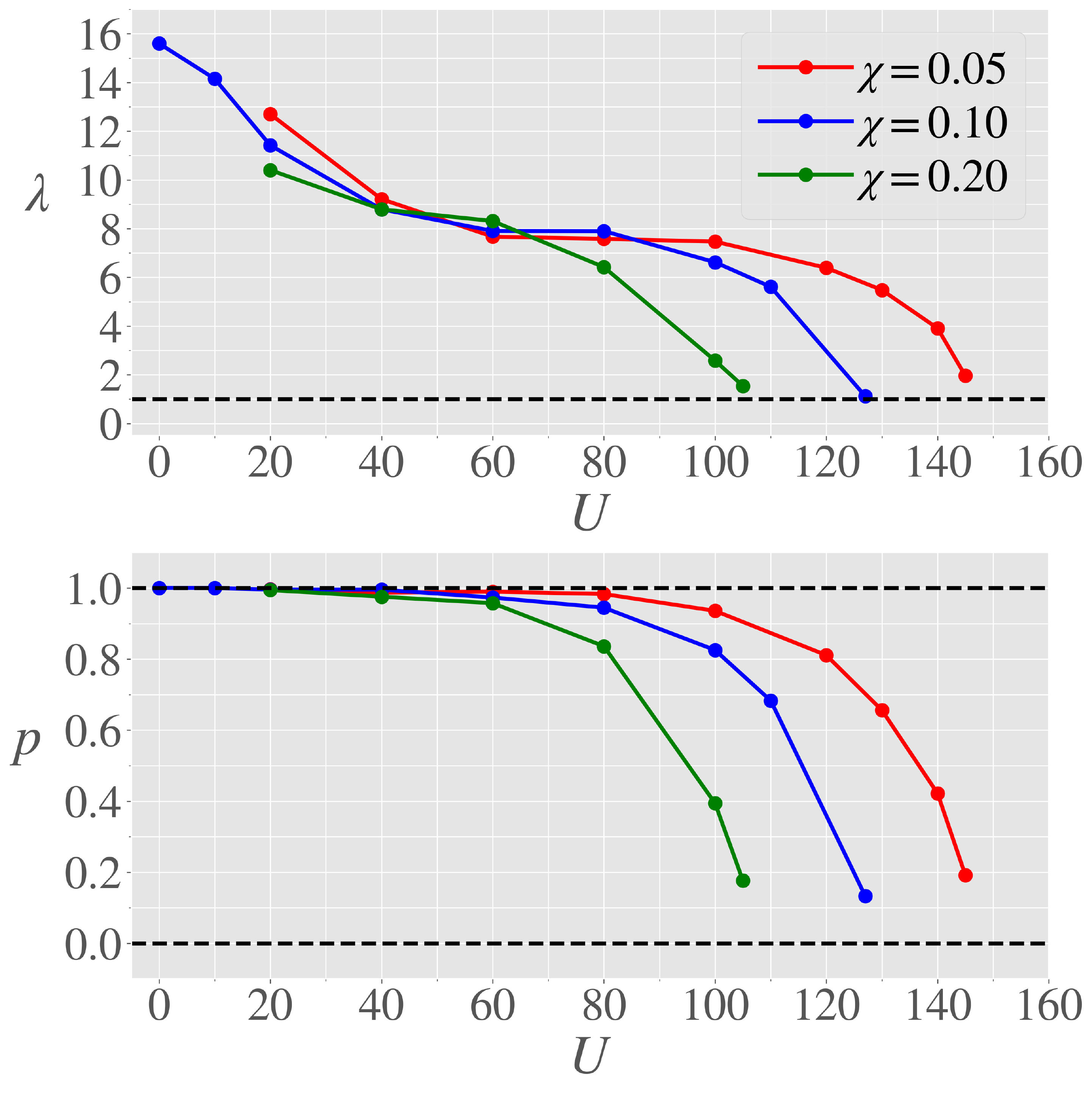}
  \caption{(Top) Order parameter $\lambda$ indicating the length of the boundary of  aggregates formed by the passive particles relative to that of an ideal hexagon structure as a function of the active velocity $U$, for different values of the  ratio $\chi$ between active and passive particles in the system, at constant volume fraction $\phi=0.2$. (Bottom) Phase ratio, $p$, of passive particles in the solid state divided by the total number of passive particles as a function of the active velocity $U$, for different values of $\chi$, at constant volume fraction $\phi=0.2$.}
 \label{figure2}
\end{figure}

\begin{figure}[t!]
 \centering
    \includegraphics[width=.475\textwidth]{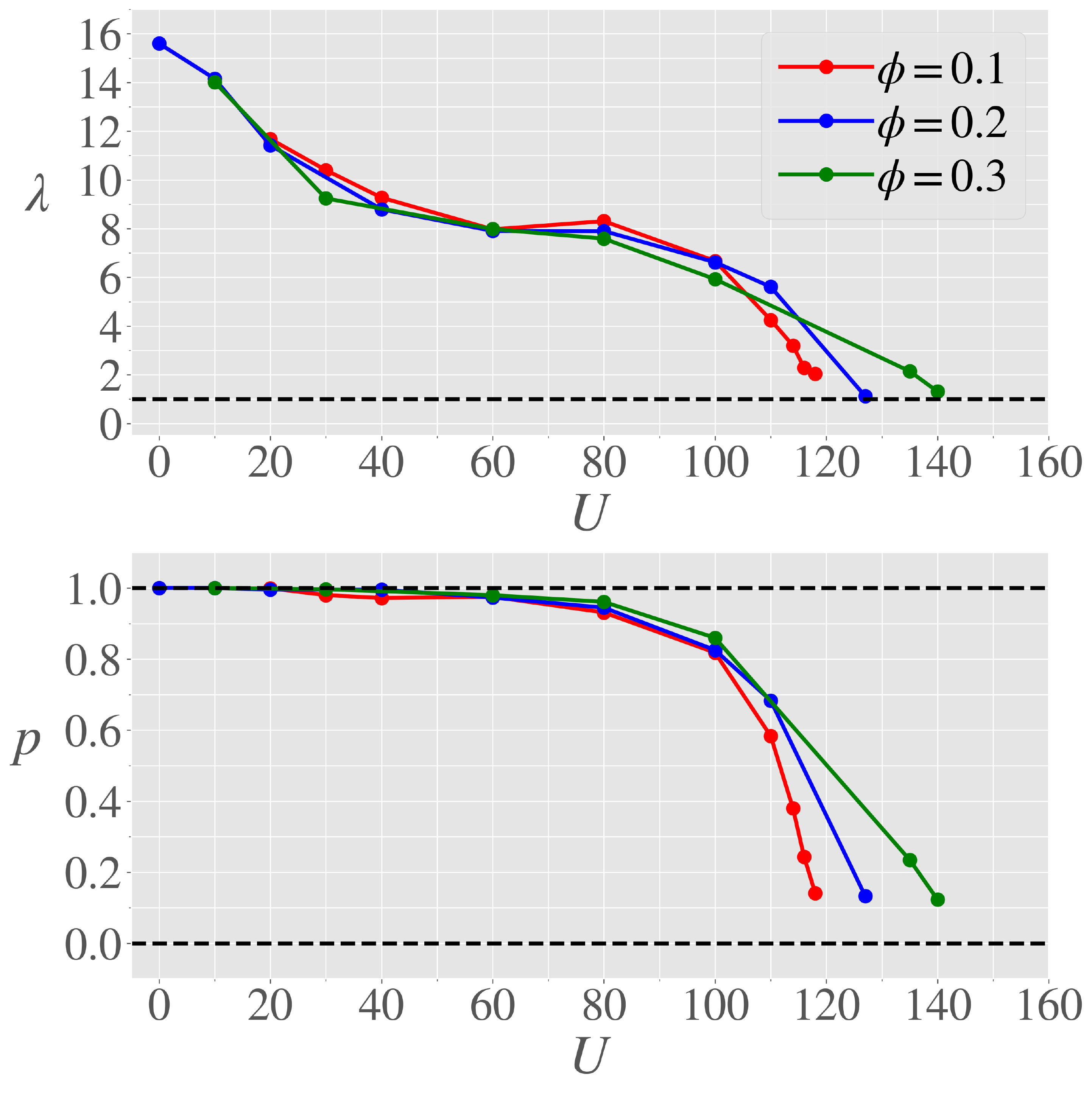}
  \caption{(Top) Order parameter $\lambda$ indicating the length of the boundary of  aggregates formed by the passive particles relative to that of an ideal hexagonal structure as a function of the active velocity $U$, for different volume fractions $\phi$ at constant $\chi=0.1$. (Top) Phase ratio, $p$, of passive particles in the solid state divided by the total number of passive particles as a function of the active velocity $U$, for different values of $\phi$ at constant $\chi=0.1$.}
 \label{figure3}
\end{figure}

The results for the isotropic case in two dimensions are summarized in Figs. 1-3.
Figure 1 highlights different snapshots of the typical steady state configurations obtained for a low density system ($\phi=\pi/4(N+M)(\sigma/L)^2$=0.2)  of isotropically attractive particles self-assembling in the presence of a 10\% fraction of  purely-repulsive active particles ($\chi=0.1$). 
The results are quite remarkable. 
Figure 1(A) shows the  extended colloidal gel formed by the passive building blocks when activity is turned off ($U=0$).
Figures 1(B-C) show how the gel is dissolved in favor of a stable cluster phase of intermediate sized crystals for activities $U\in(40-100$). The number of clusters increases and become more isotropic in shape as activity is increased.
  As a visual aid to understand the dynamics of this phase of monodispere crystallites see supplemental videos S1 and S2. 
Finally, Fig. 1(D), obtained for $U\simeq 125$ shows a single perfect crystal coexisting with its own fluid.
For even larger values of the active forces, the system melts completely into a gas, permitting just a few small and short-lived clusters. 
  The phenomenology that we observe in our system is overall consistent with the work of Omar et al. \cite{Omar2019}, where they study the dynamics of attractive spherical colloidal particles forming a colloidal gel in the presence of an attractive active dopant. However, there are some important differences regarding the stability of the clusters as we discuss below. 
For the interested reader, the authors in \cite{Omar2019} introduce a mechanical framework and various scaling arguments, that are relevant to a range of active doped systems including the ones discussed in this work.  

To quantitatively characterize the morphology of the aggregates as a function of the self-propelling speed, we compute a normalized surface to volume ratio of the passive structures. 
By considering the total perimeter of the condensed structures in the system and dividing that length by the perimeter of an ideal perfect hexagonal crystal having the same number of particles \footnote{The length of the boundary for a perfect hexagonal crystal containing $N$ particles is fairly well described by the empirical function $f(N)=-3.309229+3.47188\,N^{\frac{1}{2}}$.}.
This order parameter, $\lambda$, acquires a value much larger than one when the passive particles form an extended gel, and tends to one when in the presence of a single defect-free hexagonal crystal. 
As activity is increased, more and more passive colloids are found in the gas phase or form small dynamic crystallites of only tens of colloids.
We exclude small short-lived clusters that   contain less than ($N_{\mathrm cut} < 40$) in our analysis . 
The results are rather insensitive to the specific values of $N_{\mathrm cut}$ when taken within the range $N_{\mathrm cut}=20-60$.
The top panel of Fig.~\ref{figure2} shows the dependence of $\lambda$ on $U$ for three different ratios of active and passive particles $\chi$ at a fixed system volume fraction $\phi=0.2$. 
Whereas the top panel of Fig.~\ref{figure3} shows the dependence of $\lambda$ on $U$ for three different volume fractions $\phi$ at a fixed ratio $\chi=0.1$.
We also report in the bottom panel of Figs.~\ref{figure2} and ~\ref{figure3} the corresponding phase fraction $p$ defined as the fraction of passive particles that belong to a cluster larger than $N_{\mathrm cut}$ as a function of $U$.

As anticipated from the simulation snapshots, the overall behavior for all values of $\phi$ and $\chi$ indicates that $\lambda$ rapidly decays as soon as activity is introduced in the system.
This rapid drop, which physically corresponds to the collapse of the gel network, is followed by a plateauing region where $\lambda$ is only weakly dependent on $U$. 
This flat region in $\lambda$ corresponds to an ensemble of crystallites which become increasing more monodisperse and numerous as $U$ is increased. 
At sufficiently large $U$, $\lambda$ obtains the nearly ideal value of one, where a single or only a few crystals of passive particles are present in solution.
It is important to note how the phase fraction $p$ remains constant and nearly one until the final decay of $\lambda$. 
This indicates that for most values of $U$ every passive building block in the system belongs to a cluster of at least size $N_{cut}$, be that a part of some extended gel network or compact crystallite.
The decrease in the number of clusters which accompanies the final decay of $\lambda$ is followed by a   sharp decay in $p$ , indicating that most particles are now in highly dynamic clusters that contains less than $N_{\mathrm cut}$ particles.

\begin{figure}[t!]
 \centering
    \includegraphics[width=.475\textwidth]{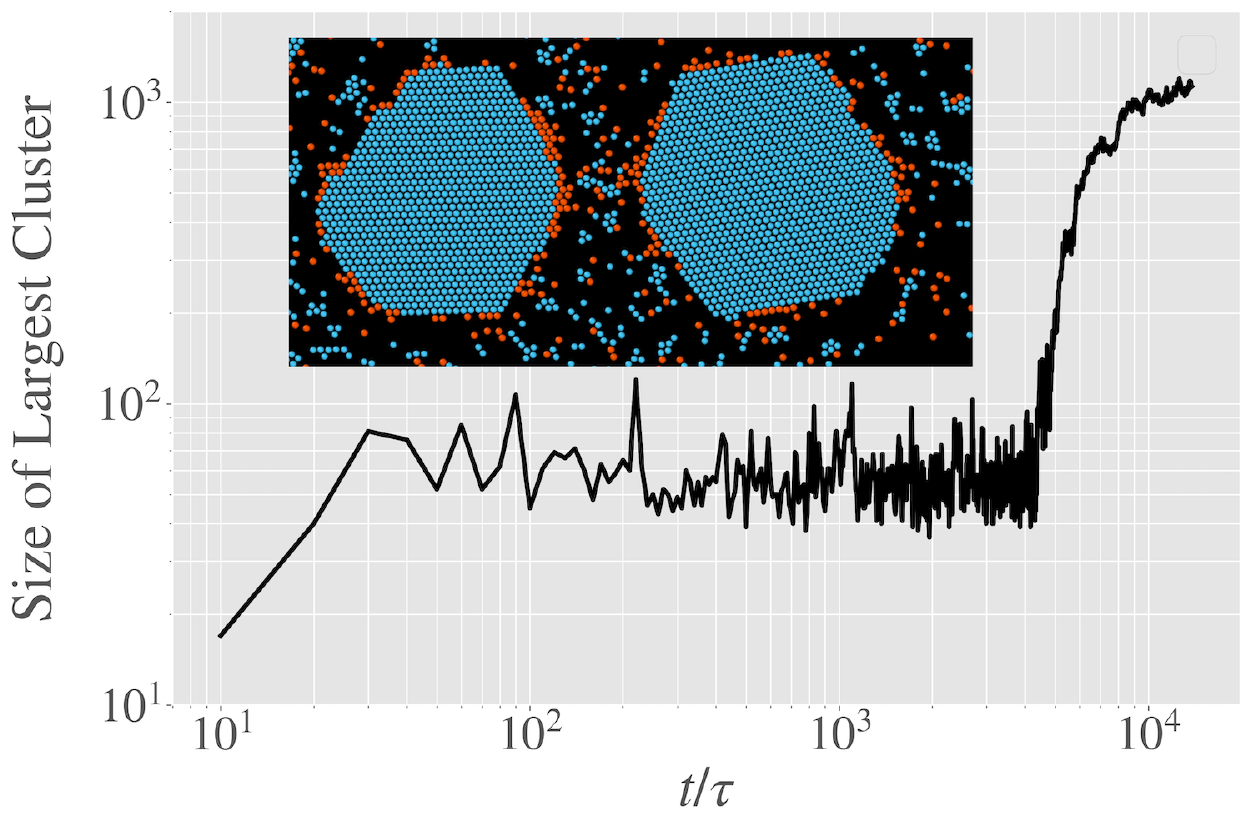}
  \caption{Log-log plot of the size of the largest cluster in the system over time for $\phi=0.2$, $\chi=0.1$ and $U=127$. Here the total number of particles is 8192.
  The inset shows typical finite size clusters of passive particles (in blue) shielded by active particles (in orange) for $\phi=0.2$, $\chi=0.2$. }
 \label{figure4}
\end{figure}

The behavior of $\lambda$ at a fixed volume fraction, as shown in Fig. 2, illustrates that larger values of $\chi$ require smaller active velocities before the system is fully fluidized.
This is expected, as more active colloids can collective exert larger forces on a particular region of the passive structure and have a higher tendency of locally reshaping it or breaking it down. 
The results at fixed $\chi$ as provided in Fig. 3 indicate that larger activities are required to drive the system into the gas phase as the volume fraction of the system is increased. 
This result is rationalized by appreciating that the larger the volume fraction, the more readily passive particles can find each other and rebind. 
Thus faster particles are required to break down the passive structures.

There are two nontrivial aspects of our results which require further investigation.
(1) The behavior of the system for intermediate values of $U$, which are characterized by an ensemble of similarly-sized stable hexagonal crystallites.
(2) The crystal growth dynamics associated with the formation of the single crystal for the largest values of $U$ prior to the complete fluidization of the system which follows what appears to be a traditional crystal nucleation pathway. 

To test whether the finite-size crystallites are not a mere transient step in a slow cluster-to-cluster association dynamic, we extend the duration of our simulations up to $10^{10}$ timesteps.
Within this extended duration of $10^5\tau$, clusters have had the opportunity to merge with one another numerous times, yet once they acquire a well defined size, their growth appears to stop.
A detailed analysis of the particle configurations reveals that once the clusters are sufficiently large, their surfaces become partially covered by a dynamic corona of mobile active colloids, and their presence effectively passivates their surface by providing a layer of hard-sphere particles that shields them against coalescence with other clusters.   The average number of active particles on the surface of the clusters depends on their overal concentration, their activity, and the size of the clusters. 
The inset in Figure~\ref{figure4} shows a pair of typical crystallites with a dynamic corona of active colloids around each. 
This is reminiscent of the addition of a passivating ligand in nanocrystal self-assembly to prevent further growth and allow stabilization of preferred crystal size.
 The supplemental video S2 highlights this shielding mechanism and the dynamic layering that takes place at the surface of a passive cluster. 
For two clusters to successfully merge, it can take a significant amount of time relative to the timescale associated with active particles colliding with a passive cluster. 
The presence of even a few active particles on the surface of the clusters is sufficient to prevent the merging and coalescence process. 
Two clusters need to merge in the proper orientation to maximize the stability of their attractive interaction.
Ideally, to form the most stable structure two cluster edges would align and come together. 
More often, we observe two clusters forming a temporary contact with a handful of particles on each cluster. 
This weak bond between clusters can easily be severed by active particles in suspension before it is able to reorganize and further coalesce. 

\begin{figure}[b!]
 \centering
    \includegraphics[width=.475\textwidth]{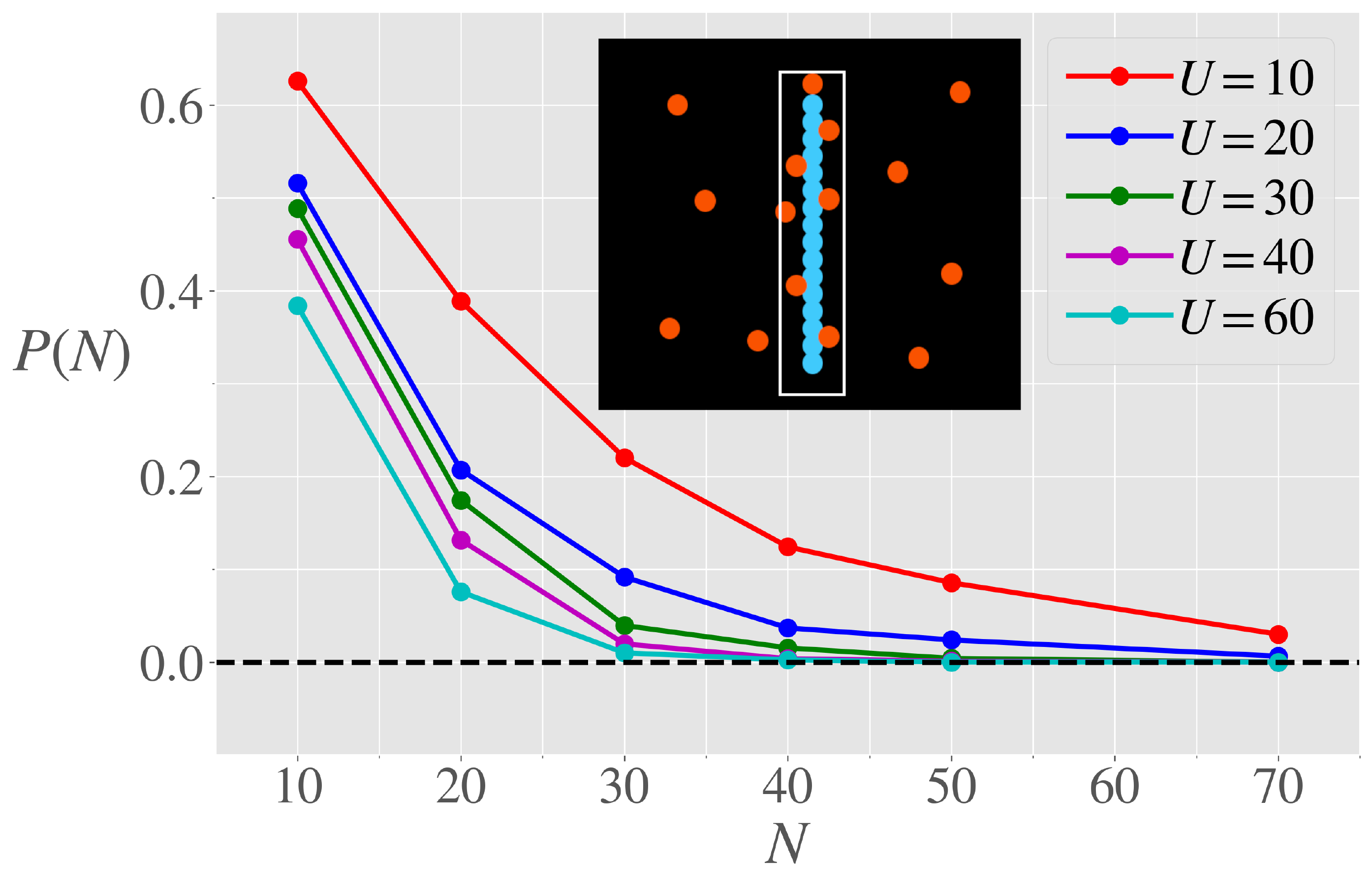}
  \caption{Probability of finding no active particle, $P(N)$, within $1.5\sigma$ from a frozen interface build by placing $N$ particles of diameter $\sigma$ next to each other in linear formation. Different curves correspond to different speeds. The volume fraction of active particles is $0.01$.  The inset sketches the geometry for $N=16$ and the probability $P(N)$ is computed based off of the box about the fixed passive structure.}  
 \label{figure5}
\end{figure}

\begin{figure*}[t!]
 \centering
    \includegraphics[width=.95\textwidth]{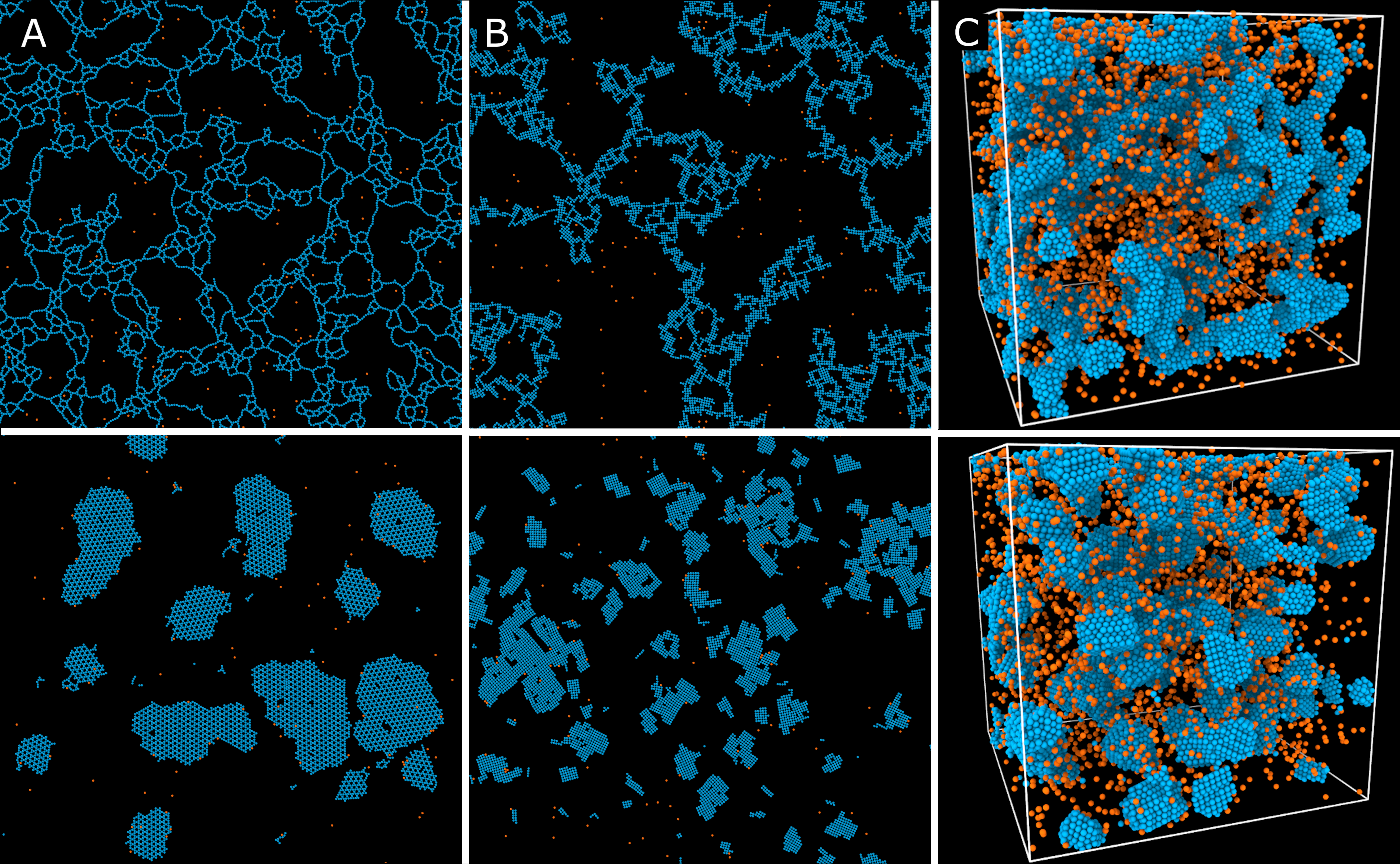}
  \caption{Additional examples of the reshaping and healing of colloidal gels with complex building blocks. The top panel of figure A and B correspond to the equilibrium structure ($U=0$) of triblock Janus particles and four-patch particles for a volume fraction of $\phi=0.2$ and active doping fraction $\chi=0.016$. The bottom panel of Fig A and B highlight the stable cluster of the kagome lattice and square symmetry crystal, respectively. Here, the active velocity is $U=50$ and $U=55$, respectively. The top panel of Fig.~\ref{figure6}C depicts the extended gel for isotropically attractive particles in three dimensions and the bottom panel shows the suspension of monodisperse crystallites for the same building block. In the three dimensions case, $\phi=0.1$, $\chi=0.1$, and $U=100$}
 \label{figure6}
\end{figure*}

Several studies  have considered the accumulation of active particles on flat and curved surfaces \cite{Elgeti2013,leite_depletion_2016,Harder2014,Fily2014,Smallenburg2015,Ray2014,Mallory2014a,Ni2015,yan2015force}.
Surface accumulation of active Brownian particles is characterized by a density profile that peaks at the surface and decays exponentially as one moves away from it. 
In most of these studies, infinitely long surfaces with non-interacting spherical active particles were considered. 
In order to capture the role of finite size effects and volume exclusion on the accumulation of active particles on the surface of a passive cluster, we performed a series of simulations where we construct a fixed surface by laying down $N$ particles next to each other at a distance of    $\sigma$ in a linear formation and add a small fraction of active particles in solution.
  A sketch of the geometry is shown in the inset of Fig. 5.  
We then measure the probability that no active particle is found in contact with the surface (i.e. within a distance   $1.5 \sigma$  from any of the fixed surface particles). 
$P(N)$ as a function of $N$, for different values of $U$ at an active particle volume fraction of   $0.01$ are shown in Fig.~\ref{figure5}. 
  This particular value was chosen as we were interested in conducting a set of simulations mimicking the conditions of the original suspension with an overall volume fraction $\phi=0.1$ with a doping fraction of $\chi=0.1$.  
The results indicate that already for moderate active velocities, $P(N)$ quickly decays to zero. 
  In other words, once a cluster reaches a characteristic size there will always be some number of active particles at the surface of the cluster, and this fraction of active particles at the surface will only increase with cluster size further passivating the surface and arresting the  coalescence of passive clusters. 

This is an important result for two reasons.
First, it suggests a simple physical mechanism to understand the stabilization of the finite size clusters  against cluster-to-cluster coalescence .
The second is that it explains the nucleation-like crystal formation observed for large activities.
In this limit, only a sufficiently large isotropic cluster can withstand the strength of the active forces.
Starting from a homogeneous uniform state, the formation of such a large cluster is rare and can take a significant amount of time (See Fig. 4).
However, once it forms, its growth can be easily fueled by the addition of either single particles, or very small clusters (for which the surface passivisation effect is weak).
The final configuration in this case is a large cluster coexisting with its own fluid.
In Fig.~\ref{figure4}, a typical time trace for the size of the largest crystal over time in a system of 8196 particle with $\chi=0.1$, at $\phi=0.2$.
This is for a value of the active velocity $U=127$ where the system exhibits nucleation-like behavior.
By inspection, the size of the critical nucleus is between 100-120 particles.
Any cluster with a size below this will melt, but once it becomes larger than that onset value it can easily grow.

  The behavior of the suspension in the region of self-propelling speeds where not all passive particles belong to a cluster, but a significant number of them remain in the fluid phase, can also be though of as a competition between passive particles attempting to join a cluster and active particles sculpting the cluster surface by removing single or small clusters of passive particles from the interface.
This is indeed the reason behind the sharply faceted nature of clusters in the system.
While this simple mechanism can explain the finite size of single clusters for values of activities near and below the onset of fluidization, the repulsion between large clusters is really due to steric and dynamic layering of active particles discussed above.

To further check for the stability of the cluster phase, we also ran a number of simulations starting from an initial configuration where all passive particles where prepared into a single large hexagonal crystal, with the active particles randomly placed in the simulation box. After a sufficiently long time, we observed that the active particles where able to slowly chip away single particles from the initial crystal, and eventually a cluster phase that persisted over time was recovered.

It is tempting to think of the small concentration of active colloids as simply generating larger temperature fluctuations in the solution.
This effective temperature mapping is only reasonable in the large activity limit where crystallization proceeds via the process of nucleation.
To confirm this, we also performed a series of simulations where the fraction of purely-repulsive active colloids are replaced with purely Brownian particles fluctuating at temperatures on the order of $\sim U^2$.
For large values of this effective temperature, we indeed observed crystal nucleation from a fluid phase.
The phenomenology is however rather different for lower effectively temperatures.
In the regime where the active system exhibits the stable microphase of long lived finite-size crystals the temperature-mapped system consistently shows continuous cluster aggregation.
In the active case, the doped particles accumulate at the surface of passive clusters and form a dynamic shield  preventing further cluster aggregation  (See supplementary videos S1 and S2) .
In the case where the doped particles are Brownian with an effective higher temperature there is no surface accumulation and these particles serve as a non-adsorbing depletant.
The net effect is an attractive depletion interaction between the clusters that favors coalescence.
What makes active particles particularly unique and is the driving factor for much of the phenomenology in this study is that their motion is correlated over a length scale $\ell=U/D_R$, which is often called the run-length and is analogous to the concept of persistence length in polymer physics. 
An understanding of the role of this length scale can be used to great effect to exhibit further control over the self-assembly process.
 For an in-depth discussion on the role of run-length in reshaping colloidal gels via active doping see \cite{Omar2019}.  

As a poignant example of this control we outline two different protocols which can be implemented to increase the size of the largest crystals once the system reaches steady-state.
These strategies are most effective when $U$ is sufficiently large such that the system is either in the phase of monodisperse crystallites or is a single cluster coexisting with its own fluid.
The first strategy is to slowly increase the rate of rotational diffusion of the active colloids, which in effect decreases their run-length.
This has the net result of reducing the accumulation of the particles on the crystal surfaces and promotes its growth.
In this limit $\ell \rightarrow 0$, the active system becomes analogous to the thermally-mapped system discussed above.
In simulation, modulating the rate of rotational diffusion of the active colloids is trivial.
However, experimentally this proves to be more challenging and is an area of research that has garnered considerable interest.
Control over the rotational degrees of freedom of active colloids has been shown to be dependent on the specific propulsion mechanism through the associated phoretic and near-field hydrodynamic interactions which can vary widely.
We find it highly intriguing that in most instances phoretic and near-field hydrodynamic effects are highly tunable suggesting another powerful handle to modulating system conditions   (See \cite{Moran2017,Popescu2018, Vutukuri2020} and references therein) .
A natural extension of this work would be to explicitly model and quantify these contributions for a variety of active colloids with different propulsion mechanisms.
The second strategy is currently more easily experimentally accessible and that is simply to slowly decrease the active speed of the particles.
This type of control over the active velocity has already been demonstrated for light-activated active colloids.
In a similar fashion to decreasing the rate of rotational diffusion, this also has the effect of decreasing the number of particles shielding the clusters promoting its grows, and simultaneously lowers the crystal nucleation barrier.
In both cases we observe that the size of the largest cluster in the system can be easily doubled or tripled. 

Perhaps the most appealing features of active doping is that the mechanism appears to be independent of the details of the interparticle interaction between passive building blocks as long as it is sufficiently short ranged.
These more sophisticated building blocks with patchy or anisotropic pair interaction allow for the self-assembly of structures of increased complexity.
To illustrate the generic nature of this approach, we considered two different self-assembly problems with patchy particles.
The first are triblock Janus colloids with an attractive patch at each pole which spontaneously self-assemble into a kagome lattice and the second are particles with four short ranged attractive patches designed to form a simple square lattice in two dimensions   (see supplemental text for details of pair interaction) .
In a similar spirit to the isotropic case, the binding energy between patches for both sets of patchy particles were chosen to be sufficiently strong such that the system rapidly evolves into a metastable extended gel configuration.
The equilibrium steady state configuration are shown in the top panels of Fig. 6A and Fig. 6B for the triblock and four-patch particles, respectively.
The active force is chosen such that it is larger than the binding energy between two patchy particles.
For the triblock and four patch system $U=50$ and $U=55$, respectively. 
This creates conditions such that a single active colloid can break the strands that characterize the gel, but is unable to destroy larger aggregates as they form and can only reshape their surface.
The volume fraction for both systems here is $\phi=0.2$ and the doping fraction is $\chi=0.016$.
The phenomenology for these patchy particle system follows the same trend as laid out for the isotropic case.
As activity is increased, the gel collapses and the system is driven into a configuration of finite crystallites.
  See supplemental videos S3-S6 for typical behavior of the triblock Janus and four-patch particle systems at equilbrium and in the presence of an active dopant .
A snapshot of the microphase of monodisperse crystallites is shown in the bottom panels of Fig. 6A and Fig. 6B for the triblock and four-patch particles, respectively.
These crystals can be made more numerous and regular in shape by further increasing the active velocity.
Both of these systems are amenable to the protocols laid out in the previous section to increase the size of the clusters. 
Additionally, if activity is further increased, both systems will be fluidized and it will only be possible to form small transient clusters.
Similar phenomenology and phase behavior is also observed for building blocks interacting through an isotropic potential in three dimensions, as shown in Fig.~\ref{figure6}C which depicts the extended gel in the top panel and the suspension of monodisperse crystallites in the bottom panel. 
 
The building blocks here interact through the same Mie potential as the two dimensional case.
In the three dimensions case shown in in Fig.~\ref{figure6}C the system parameters are: $\phi=0.1$, $\chi=0.1$, and $U=100$.
Lastly, we have predominately considered building blocks with a fairly short range interaction which promotes the formation of kinetically frustrated aggregates, we have also performed a series of simulations with a (12-6) Lennard-Jones potential with a cut-off at $2.5\sigma$, and found that an analogous mechanism of stabilization exists even in this case with a longer range interaction. 

\section{Conclusion}

Our results indicate that it is always possible to add a sufficient number of purely-repulsive active colloids at the appropriate self-propelling speed to break apart any arrested colloidal gels and allow the exclusive and controlled formation of isotropic, ordered structures.
 It is important to emphasize that whether we implement the active doping strategy once the colloidal gel has already formed or from a random fluid state of the passive particles, the final results are the same.
In systems where activity can be dynamically modulated, our results suggest a number of protocols that can be implemented to maximize self-assembly.
One can think of self-propulsion in these systems as a very selective filter that only allows for the stabilization of certain structures. 
This is possible because the persistent active forces can rapidly collapse and reshape the elongated linear strands making up the gel network into large isotropic clusters.
Remarkably, this mechanism is independent of the specific pair interaction between the passive building blocks.
Interestingly for all cases considered, it is possible to stabilize a suspension of fairly monodisperse finite-size crystals for a range of intermediate self-propelling velocities.
The overall size of the clusters decreases with an increase in strength of the active forces, while their overall number increases, until, beyond an onset value of the active force the system completely melts into a fluid.
We study in detail the mechanism by which these finite size clusters are stabilized and discovered two important contribution which promote their stability.
The first is a steric stabilization due to the presence of a layer of active colloids at the boundary of each crystallite.
This dynamic but persistent layer prevents individual clusters from coming in contact and merging.
The second contribution is related to the emergence of the swim pressure \cite{Takatori2014,Mallory2014b,Fily2012} as the clusters grow to a size comparable or much larger than the persistence length of an active colloid.
Once the clusters have achieved this sufficiently large size, there is a statistical force, similar in spirit to depletion in equilibrium system, which is extremely repulsive at short interaction distances and weakly attractive at longer distances.
This effect in active matter systems has been studied in some detail elsewhere \cite{leite_depletion_2016,Harder2014,Ray2014,ZaeifiYamchi2017}.
The active stabilization of these clusters is quite remarkable and cannot be described by mapping the active forces into an effective temperature of the solution.
Furthermore, the mechanism of stabilization is significantly different than the one proposed for the stabilization of ``living" crystals in   suspensions of attractive active colloids  ~\cite{mognetti_living_2013}.

 Lastly, it should be mentioned that in this work we neglect all hydrodynamic and phoretic interactions between colloids. Such details may very well, and probably will, affect the efficacy of the active doping approach. 
At these early stages in the study of active doping, we preferred not to specify a particular propulsion mechanism and instead  considered active colloids that self-propel at a fixed speed. Although we studied this problem through the lens of a minimal model for the propulsion mechanism, we believe the general approach and methodology discussed here can be applied to a large class of colloidal self-assembly problems. We are optimistic that phoretic and hydrodynamic effects will offer another powerful handle to modulating system conditions. 

\section*{Supplementary Material}

See supplementary material for details on the patchy particle pair potential and movies S1-S6 which illustrate various phenomenological aspects of the active doping process for both isotropic and patchy passive building blocks. 

\begin{acknowledgements}
A.C. acknowledges financial supported from the National Science Foundation under Grant No. DMR-1703873.  S.A.M. acknowledges financial support from the Arnold and Mabel Beckman Foundation. We gratefully acknowledge the support of the NVIDIA Corporation for the donation of the Titan V GPU used to carry out this work. M.L.B. acknowledges support from Columbia University through the Guthikonda Fellowship. We thank Austin Dulaney, Ahmad Omar and Hyeongjoo Row for insightful discussions and a critical reading of an early version of the manuscript. The authors declare no competing financial interest.
\end{acknowledgements}

\section*{Data Availability Statement}

The data that support the findings of this study are available from the corresponding author upon reasonable request.
%\bibliography{references}

\begin{thebibliography}{95}%
\makeatletter
\providecommand \@ifxundefined [1]{%
 \@ifx{#1\undefined}
}%
\providecommand \@ifnum [1]{%
 \ifnum #1\expandafter \@firstoftwo
 \else \expandafter \@secondoftwo
 \fi
}%
\providecommand \@ifx [1]{%
 \ifx #1\expandafter \@firstoftwo
 \else \expandafter \@secondoftwo
 \fi
}%
\providecommand \natexlab [1]{#1}%
\providecommand \enquote  [1]{``#1''}%
\providecommand \bibnamefont  [1]{#1}%
\providecommand \bibfnamefont [1]{#1}%
\providecommand \citenamefont [1]{#1}%
\providecommand \href@noop [0]{\@secondoftwo}%
\providecommand \href [0]{\begingroup \@sanitize@url \@href}%
\providecommand \@href[1]{\@@startlink{#1}\@@href}%
\providecommand \@@href[1]{\endgroup#1\@@endlink}%
\providecommand \@sanitize@url [0]{\catcode `\\12\catcode `\$12\catcode
  `\&12\catcode `\#12\catcode `\^12\catcode `\_12\catcode `\%12\relax}%
\providecommand \@@startlink[1]{}%
\providecommand \@@endlink[0]{}%
\providecommand \url  [0]{\begingroup\@sanitize@url \@url }%
\providecommand \@url [1]{\endgroup\@href {#1}{\urlprefix }}%
\providecommand \urlprefix  [0]{URL }%
\providecommand \Eprint [0]{\href }%
\providecommand \doibase [0]{https://doi.org/}%
\providecommand \selectlanguage [0]{\@gobble}%
\providecommand \bibinfo  [0]{\@secondoftwo}%
\providecommand \bibfield  [0]{\@secondoftwo}%
\providecommand \translation [1]{[#1]}%
\providecommand \BibitemOpen [0]{}%
\providecommand \bibitemStop [0]{}%
\providecommand \bibitemNoStop [0]{.\EOS\space}%
\providecommand \EOS [0]{\spacefactor3000\relax}%
\providecommand \BibitemShut  [1]{\csname bibitem#1\endcsname}%
\let\auto@bib@innerbib\@empty
%</preamble>
\bibitem [{\citenamefont {Glotzer}\ \emph {et~al.}(2004)\citenamefont
  {Glotzer}, \citenamefont {Solomon},\ and\ \citenamefont
  {Kotov}}]{Glotzer2004a}%
  \BibitemOpen
  \bibfield  {author} {\bibinfo {author} {\bibfnamefont {S.~C.}\ \bibnamefont
  {Glotzer}}, \bibinfo {author} {\bibfnamefont {M.~J.}\ \bibnamefont
  {Solomon}},\ and\ \bibinfo {author} {\bibfnamefont {N.~A.}\ \bibnamefont
  {Kotov}},\ }\bibfield  {title} {\bibinfo {title} {{Self-assembly: From
  nanoscale to microscale colloids}},\ }\href@noop {} {\bibfield  {journal}
  {\bibinfo  {journal} {AIChE J.}\ }\textbf {\bibinfo {volume} {50}},\ \bibinfo
  {pages} {2978} (\bibinfo {year} {2004})}\BibitemShut {NoStop}%
\bibitem [{\citenamefont {Glotzer}(2004)}]{glotzer_assembly_2004}%
  \BibitemOpen
  \bibfield  {author} {\bibinfo {author} {\bibfnamefont {S.~C.}\ \bibnamefont
  {Glotzer}},\ }\bibfield  {title} {\bibinfo {title} {Some assembly required},\
  }\href@noop {} {\bibfield  {journal} {\bibinfo  {journal} {Science}\ }\textbf
  {\bibinfo {volume} {306}},\ \bibinfo {pages} {419} (\bibinfo {year}
  {2004})}\BibitemShut {NoStop}%
\bibitem [{\citenamefont {Sacanna}\ \emph {et~al.}(2013)\citenamefont
  {Sacanna}, \citenamefont {Pine},\ and\ \citenamefont {Yi}}]{Sacanna2013a}%
  \BibitemOpen
  \bibfield  {author} {\bibinfo {author} {\bibfnamefont {S.}~\bibnamefont
  {Sacanna}}, \bibinfo {author} {\bibfnamefont {D.~J.}\ \bibnamefont {Pine}},\
  and\ \bibinfo {author} {\bibfnamefont {G.~R.}\ \bibnamefont {Yi}},\
  }\bibfield  {title} {\bibinfo {title} {{Engineering shape: The novel
  geometries of colloidal self-assembly}},\ }\href@noop {} {\bibfield
  {journal} {\bibinfo  {journal} {Soft Matter}\ }\textbf {\bibinfo {volume}
  {9}},\ \bibinfo {pages} {8096} (\bibinfo {year} {2013})}\BibitemShut
  {NoStop}%
\bibitem [{\citenamefont {Li}\ \emph {et~al.}(2011)\citenamefont {Li},
  \citenamefont {Josephson},\ and\ \citenamefont {Stein}}]{Li2011}%
  \BibitemOpen
  \bibfield  {author} {\bibinfo {author} {\bibfnamefont {F.}~\bibnamefont
  {Li}}, \bibinfo {author} {\bibfnamefont {D.~P.}\ \bibnamefont {Josephson}},\
  and\ \bibinfo {author} {\bibfnamefont {A.}~\bibnamefont {Stein}},\ }\bibfield
   {title} {\bibinfo {title} {{Colloidal assembly: The road from particles to
  colloidal molecules and crystals}},\ }\href@noop {} {\bibfield  {journal}
  {\bibinfo  {journal} {Angew. Chemie - Int. Ed.}\ }\textbf {\bibinfo {volume}
  {50}},\ \bibinfo {pages} {360} (\bibinfo {year} {2011})}\BibitemShut
  {NoStop}%
\bibitem [{\citenamefont {Sacanna}\ and\ \citenamefont
  {Pine}(2011)}]{Sacanna2011}%
  \BibitemOpen
  \bibfield  {author} {\bibinfo {author} {\bibfnamefont {S.}~\bibnamefont
  {Sacanna}}\ and\ \bibinfo {author} {\bibfnamefont {D.~J.}\ \bibnamefont
  {Pine}},\ }\bibfield  {title} {\bibinfo {title} {{Shape-anisotropic colloids:
  Building blocks for complex assemblies}},\ }\href@noop {} {\bibfield
  {journal} {\bibinfo  {journal} {Curr. Opin. Colloid Interface Sci.}\ }\textbf
  {\bibinfo {volume} {16}},\ \bibinfo {pages} {96} (\bibinfo {year}
  {2011})}\BibitemShut {NoStop}%
\bibitem [{\citenamefont {Zhang}\ \emph {et~al.}(2015)\citenamefont {Zhang},
  \citenamefont {Luijten},\ and\ \citenamefont {Granick}}]{Zhang2015}%
  \BibitemOpen
  \bibfield  {author} {\bibinfo {author} {\bibfnamefont {J.}~\bibnamefont
  {Zhang}}, \bibinfo {author} {\bibfnamefont {E.}~\bibnamefont {Luijten}},\
  and\ \bibinfo {author} {\bibfnamefont {S.}~\bibnamefont {Granick}},\
  }\bibfield  {title} {\bibinfo {title} {{Toward Design Rules of Directional
  Janus Colloidal Assembly}},\ }\href
  {http://www.annualreviews.org/doi/10.1146/annurev-physchem-040214-121241}
  {\bibfield  {journal} {\bibinfo  {journal} {Annu. Rev. Phys. Chem.}\ }\textbf
  {\bibinfo {volume} {66}},\ \bibinfo {pages} {581} (\bibinfo {year}
  {2015})}\BibitemShut {NoStop}%
\bibitem [{\citenamefont {Jankowski}\ and\ \citenamefont
  {{C. Glotzer}}(2012)}]{Jankowski2012}%
  \BibitemOpen
  \bibfield  {author} {\bibinfo {author} {\bibfnamefont {E.}~\bibnamefont
  {Jankowski}}\ and\ \bibinfo {author} {\bibfnamefont {S.}~\bibnamefont
  {{C. Glotzer}}},\ }\bibfield  {title} {\bibinfo {title} {{Screening and
  designing patchy particles for optimized self-assembly propensity through
  assembly pathway engineering}},\ }\href
  {http://pubs.rsc.org/en/Content/ArticleLanding/2012/SM/C2SM07101K
  http://pubs.rsc.org/en/Content/ArticlePDF/2012/SM/C2SM07101K
  http://pubs.rsc.org/en/content/articlehtml/2012/sm/c2sm07101k} {\bibfield
  {journal} {\bibinfo  {journal} {Soft Matter}\ }\textbf {\bibinfo {volume}
  {8}},\ \bibinfo {pages} {2852} (\bibinfo {year} {2012})}\BibitemShut
  {NoStop}%
\bibitem [{\citenamefont {Zhang}\ \emph {et~al.}(2017)\citenamefont {Zhang},
  \citenamefont {Grzybowski},\ and\ \citenamefont {Granick}}]{Zhang2017}%
  \BibitemOpen
  \bibfield  {author} {\bibinfo {author} {\bibfnamefont {J.}~\bibnamefont
  {Zhang}}, \bibinfo {author} {\bibfnamefont {B.~A.}\ \bibnamefont
  {Grzybowski}},\ and\ \bibinfo {author} {\bibfnamefont {S.}~\bibnamefont
  {Granick}},\ }\bibfield  {title} {\bibinfo {title} {{Janus Particle
  Synthesis, Assembly, and Application}},\ }\href@noop {} {\bibfield  {journal}
  {\bibinfo  {journal} {Langmuir}\ }\textbf {\bibinfo {volume} {33}},\ \bibinfo
  {pages} {6964} (\bibinfo {year} {2017})}\BibitemShut {NoStop}%
\bibitem [{\citenamefont {Ravaine}\ and\ \citenamefont
  {Duguet}(2017)}]{Ravaine2017}%
  \BibitemOpen
  \bibfield  {author} {\bibinfo {author} {\bibfnamefont {S.}~\bibnamefont
  {Ravaine}}\ and\ \bibinfo {author} {\bibfnamefont {E.}~\bibnamefont
  {Duguet}},\ }\bibfield  {title} {\bibinfo {title} {Synthesis and assembly of
  patchy particles: Recent progress and future prospects},\ }\href@noop {}
  {\bibfield  {journal} {\bibinfo  {journal} {Curr. Opin. Colloid Interface
  Sci.}\ }\textbf {\bibinfo {volume} {30}},\ \bibinfo {pages} {45} (\bibinfo
  {year} {2017})}\BibitemShut {NoStop}%
\bibitem [{\citenamefont {Morphew}\ \emph {et~al.}(2018)\citenamefont
  {Morphew}, \citenamefont {Shaw}, \citenamefont {Avins},\ and\ \citenamefont
  {Chakrabarti}}]{Morphew2018}%
  \BibitemOpen
  \bibfield  {author} {\bibinfo {author} {\bibfnamefont {D.}~\bibnamefont
  {Morphew}}, \bibinfo {author} {\bibfnamefont {J.}~\bibnamefont {Shaw}},
  \bibinfo {author} {\bibfnamefont {C.}~\bibnamefont {Avins}},\ and\ \bibinfo
  {author} {\bibfnamefont {D.}~\bibnamefont {Chakrabarti}},\ }\bibfield
  {title} {\bibinfo {title} {Programming hierarchical self-assembly of patchy
  particles into colloidal crystals via colloidal molecules},\ }\href@noop {}
  {\bibfield  {journal} {\bibinfo  {journal} {ACS nano}\ }\textbf {\bibinfo
  {volume} {12}},\ \bibinfo {pages} {2355} (\bibinfo {year}
  {2018})}\BibitemShut {NoStop}%
\bibitem [{\citenamefont {Li}\ \emph {et~al.}(2020)\citenamefont {Li},
  \citenamefont {Palis}, \citenamefont {Mérindol}, \citenamefont {Majimel},
  \citenamefont {Ravaine},\ and\ \citenamefont {Duguet}}]{Li2020}%
  \BibitemOpen
  \bibfield  {author} {\bibinfo {author} {\bibfnamefont {W.}~\bibnamefont
  {Li}}, \bibinfo {author} {\bibfnamefont {H.}~\bibnamefont {Palis}}, \bibinfo
  {author} {\bibfnamefont {R.}~\bibnamefont {Mérindol}}, \bibinfo {author}
  {\bibfnamefont {J.}~\bibnamefont {Majimel}}, \bibinfo {author} {\bibfnamefont
  {S.}~\bibnamefont {Ravaine}},\ and\ \bibinfo {author} {\bibfnamefont
  {E.}~\bibnamefont {Duguet}},\ }\bibfield  {title} {\bibinfo {title}
  {Colloidal molecules and patchy particles: complementary concepts{,}
  synthesis and self-assembly},\ }\href {https://doi.org/10.1039/C9CS00804G}
  {\bibfield  {journal} {\bibinfo  {journal} {Chem. Soc. Rev.}\ }\textbf
  {\bibinfo {volume} {49}},\ \bibinfo {pages} {1955} (\bibinfo {year}
  {2020})}\BibitemShut {NoStop}%
\bibitem [{\citenamefont {Ashton}\ \emph {et~al.}(2013)\citenamefont {Ashton},
  \citenamefont {Jack},\ and\ \citenamefont {Wilding}}]{Ashton2013}%
  \BibitemOpen
  \bibfield  {author} {\bibinfo {author} {\bibfnamefont {D.~J.}\ \bibnamefont
  {Ashton}}, \bibinfo {author} {\bibfnamefont {R.~L.}\ \bibnamefont {Jack}},\
  and\ \bibinfo {author} {\bibfnamefont {N.~B.}\ \bibnamefont {Wilding}},\
  }\bibfield  {title} {\bibinfo {title} {{Self-assembly of colloidal polymers
  via depletion-mediated lock and key binding}},\ }\href@noop {} {\bibfield
  {journal} {\bibinfo  {journal} {Soft Matter}\ }\textbf {\bibinfo {volume}
  {9}},\ \bibinfo {pages} {9661} (\bibinfo {year} {2013})}\BibitemShut
  {NoStop}%
\bibitem [{\citenamefont {Sacanna}\ \emph {et~al.}(2010)\citenamefont
  {Sacanna}, \citenamefont {Irvine}, \citenamefont {Chaikin},\ and\
  \citenamefont {Pine}}]{Sacanna2010a}%
  \BibitemOpen
  \bibfield  {author} {\bibinfo {author} {\bibfnamefont {S.}~\bibnamefont
  {Sacanna}}, \bibinfo {author} {\bibfnamefont {W.~T.}\ \bibnamefont {Irvine}},
  \bibinfo {author} {\bibfnamefont {P.~M.}\ \bibnamefont {Chaikin}},\ and\
  \bibinfo {author} {\bibfnamefont {D.~J.}\ \bibnamefont {Pine}},\ }\bibfield
  {title} {\bibinfo {title} {{Lock and key colloids}},\ }\href
  {http://www.nature.com/nature/journal/v464/n7288/pdf/nature08906.pdf
  http://www.nature.com/nature/journal/v464/n7288/abs/nature08906.html}
  {\bibfield  {journal} {\bibinfo  {journal} {Nature}\ }\textbf {\bibinfo
  {volume} {464}},\ \bibinfo {pages} {575} (\bibinfo {year}
  {2010})}\BibitemShut {NoStop}%
\bibitem [{\citenamefont {Wang}\ \emph {et~al.}(2014)\citenamefont {Wang},
  \citenamefont {Wang}, \citenamefont {Zheng}, \citenamefont {Yi},
  \citenamefont {Sacanna}, \citenamefont {Pine},\ and\ \citenamefont
  {Weck}}]{Wang2014}%
  \BibitemOpen
  \bibfield  {author} {\bibinfo {author} {\bibfnamefont {Y.}~\bibnamefont
  {Wang}}, \bibinfo {author} {\bibfnamefont {Y.}~\bibnamefont {Wang}}, \bibinfo
  {author} {\bibfnamefont {X.}~\bibnamefont {Zheng}}, \bibinfo {author}
  {\bibfnamefont {G.~R.}\ \bibnamefont {Yi}}, \bibinfo {author} {\bibfnamefont
  {S.}~\bibnamefont {Sacanna}}, \bibinfo {author} {\bibfnamefont {D.~J.}\
  \bibnamefont {Pine}},\ and\ \bibinfo {author} {\bibfnamefont
  {M.}~\bibnamefont {Weck}},\ }\bibfield  {title} {\bibinfo {title}
  {{Three-dimensional lock and key colloids}},\ }\href@noop {} {\bibfield
  {journal} {\bibinfo  {journal} {J. Am. Chem. Soc.}\ }\textbf {\bibinfo
  {volume} {136}},\ \bibinfo {pages} {6866} (\bibinfo {year}
  {2014})}\BibitemShut {NoStop}%
\bibitem [{\citenamefont {Wang}\ \emph {et~al.}(2012)\citenamefont {Wang},
  \citenamefont {Wang}, \citenamefont {Breed}, \citenamefont {Manoharan},
  \citenamefont {Feng}, \citenamefont {Hollingsworth}, \citenamefont {Weck},\
  and\ \citenamefont {Pine}}]{Wang2012a}%
  \BibitemOpen
  \bibfield  {author} {\bibinfo {author} {\bibfnamefont {Y.}~\bibnamefont
  {Wang}}, \bibinfo {author} {\bibfnamefont {Y.}~\bibnamefont {Wang}}, \bibinfo
  {author} {\bibfnamefont {D.~R.}\ \bibnamefont {Breed}}, \bibinfo {author}
  {\bibfnamefont {V.~N.}\ \bibnamefont {Manoharan}}, \bibinfo {author}
  {\bibfnamefont {L.}~\bibnamefont {Feng}}, \bibinfo {author} {\bibfnamefont
  {A.~D.}\ \bibnamefont {Hollingsworth}}, \bibinfo {author} {\bibfnamefont
  {M.}~\bibnamefont {Weck}},\ and\ \bibinfo {author} {\bibfnamefont {D.~J.}\
  \bibnamefont {Pine}},\ }\bibfield  {title} {\bibinfo {title} {{Colloids with
  valence and specific directional bonding}},\ }\href@noop {} {\bibfield
  {journal} {\bibinfo  {journal} {Nature}\ }\textbf {\bibinfo {volume} {491}},\
  \bibinfo {pages} {51} (\bibinfo {year} {2012})}\BibitemShut {NoStop}%
\bibitem [{\citenamefont {Biancaniello}\ \emph {et~al.}(2005)\citenamefont
  {Biancaniello}, \citenamefont {Kim},\ and\ \citenamefont
  {Crocker}}]{Biancaniello2005}%
  \BibitemOpen
  \bibfield  {author} {\bibinfo {author} {\bibfnamefont {P.~L.}\ \bibnamefont
  {Biancaniello}}, \bibinfo {author} {\bibfnamefont {A.~J.}\ \bibnamefont
  {Kim}},\ and\ \bibinfo {author} {\bibfnamefont {J.~C.}\ \bibnamefont
  {Crocker}},\ }\bibfield  {title} {\bibinfo {title} {{Colloidal interactions
  and self-assembly using DNA hybridization}},\ }\href@noop {} {\bibfield
  {journal} {\bibinfo  {journal} {Phys. Rev. Lett.}\ }\textbf {\bibinfo
  {volume} {94}},\ \bibinfo {pages} {58302} (\bibinfo {year}
  {2005})}\BibitemShut {NoStop}%
\bibitem [{\citenamefont {Nykypanchuk}\ \emph {et~al.}(2008)\citenamefont
  {Nykypanchuk}, \citenamefont {Maye}, \citenamefont {{Van Der Lelie}},\ and\
  \citenamefont {Gang}}]{Nykypanchuk2008}%
  \BibitemOpen
  \bibfield  {author} {\bibinfo {author} {\bibfnamefont {D.}~\bibnamefont
  {Nykypanchuk}}, \bibinfo {author} {\bibfnamefont {M.~M.}\ \bibnamefont
  {Maye}}, \bibinfo {author} {\bibfnamefont {D.}~\bibnamefont {{Van Der
  Lelie}}},\ and\ \bibinfo {author} {\bibfnamefont {O.}~\bibnamefont {Gang}},\
  }\bibfield  {title} {\bibinfo {title} {{DNA-guided crystallization of
  colloidal nanoparticles}},\ }\href@noop {} {\bibfield  {journal} {\bibinfo
  {journal} {Nature}\ }\textbf {\bibinfo {volume} {451}},\ \bibinfo {pages}
  {549} (\bibinfo {year} {2008})}\BibitemShut {NoStop}%
\bibitem [{\citenamefont {Valignat}\ \emph {et~al.}(2005)\citenamefont
  {Valignat}, \citenamefont {Theodoly}, \citenamefont {Crocker}, \citenamefont
  {Russel},\ and\ \citenamefont {Chaikin}}]{Valignat2005}%
  \BibitemOpen
  \bibfield  {author} {\bibinfo {author} {\bibfnamefont {M.-P.}\ \bibnamefont
  {Valignat}}, \bibinfo {author} {\bibfnamefont {O.}~\bibnamefont {Theodoly}},
  \bibinfo {author} {\bibfnamefont {J.~C.}\ \bibnamefont {Crocker}}, \bibinfo
  {author} {\bibfnamefont {W.~B.}\ \bibnamefont {Russel}},\ and\ \bibinfo
  {author} {\bibfnamefont {P.~M.}\ \bibnamefont {Chaikin}},\ }\bibfield
  {title} {\bibinfo {title} {{Reversible self-assembly and directed assembly of
  DNA-linked micrometer-sized colloids}},\ }\href
  {http://www.pnas.org/cgi/doi/10.1073/pnas.0500507102} {\bibfield  {journal}
  {\bibinfo  {journal} {Proc. Natl. Acad. Sci.}\ }\textbf {\bibinfo {volume}
  {102}},\ \bibinfo {pages} {4225} (\bibinfo {year} {2005})}\BibitemShut
  {NoStop}%
\bibitem [{\citenamefont {Rogers}\ \emph {et~al.}(2016)\citenamefont {Rogers},
  \citenamefont {Shih},\ and\ \citenamefont {Manoharan}}]{Rogers2016}%
  \BibitemOpen
  \bibfield  {author} {\bibinfo {author} {\bibfnamefont {W.~B.}\ \bibnamefont
  {Rogers}}, \bibinfo {author} {\bibfnamefont {W.~M.}\ \bibnamefont {Shih}},\
  and\ \bibinfo {author} {\bibfnamefont {V.~N.}\ \bibnamefont {Manoharan}},\
  }\bibfield  {title} {\bibinfo {title} {Using dna to program the self-assembly
  of colloidal nanoparticles and microparticles},\ }\href@noop {} {\bibfield
  {journal} {\bibinfo  {journal} {Nat. Rev. Mater.}\ }\textbf {\bibinfo
  {volume} {1}},\ \bibinfo {pages} {1} (\bibinfo {year} {2016})}\BibitemShut
  {NoStop}%
\bibitem [{\citenamefont {Whitelam}\ \emph {et~al.}(2009)\citenamefont
  {Whitelam}, \citenamefont {Feng}, \citenamefont {Hagan},\ and\ \citenamefont
  {Geissler}}]{whitelam_role_2009}%
  \BibitemOpen
  \bibfield  {author} {\bibinfo {author} {\bibfnamefont {S.}~\bibnamefont
  {Whitelam}}, \bibinfo {author} {\bibfnamefont {E.~H.}\ \bibnamefont {Feng}},
  \bibinfo {author} {\bibfnamefont {M.~F.}\ \bibnamefont {Hagan}},\ and\
  \bibinfo {author} {\bibfnamefont {P.~L.}\ \bibnamefont {Geissler}},\
  }\bibfield  {title} {\bibinfo {title} {{The role of collective motion in
  examples of coarsening and self-assembly}},\ }\href
  {http://pubs.rsc.org/en/Content/ArticleLanding/2009/SM/B810031D} {\bibfield
  {journal} {\bibinfo  {journal} {Soft Matter}\ }\textbf {\bibinfo {volume}
  {5}},\ \bibinfo {pages} {1251} (\bibinfo {year} {2009})}\BibitemShut
  {NoStop}%
\bibitem [{\citenamefont {Whitelam}\ and\ \citenamefont
  {Jack}(2014)}]{whitelam_statistical_2015}%
  \BibitemOpen
  \bibfield  {author} {\bibinfo {author} {\bibfnamefont {S.}~\bibnamefont
  {Whitelam}}\ and\ \bibinfo {author} {\bibfnamefont {R.~L.}\ \bibnamefont
  {Jack}},\ }\bibfield  {title} {\bibinfo {title} {{The Statistical Mechanics
  of Dynamic Pathways to Self-assembly}},\ }\href
  {http://arxiv.org/abs/1407.2505{\%}0Ahttp://dx.doi.org/10.1146/annurev-physchem-040214-121215}
  {\bibfield  {journal} {\bibinfo  {journal} {Annu. Rev. Phys. Chem.}\ }\textbf
  {\bibinfo {volume} {66}},\ \bibinfo {pages} {143} (\bibinfo {year}
  {2014})}\BibitemShut {NoStop}%
\bibitem [{\citenamefont {Miller}\ and\ \citenamefont
  {Cacciuto}(2010)}]{noauthor_exploiting_2010}%
  \BibitemOpen
  \bibfield  {author} {\bibinfo {author} {\bibfnamefont {W.~L.}\ \bibnamefont
  {Miller}}\ and\ \bibinfo {author} {\bibfnamefont {A.}~\bibnamefont
  {Cacciuto}},\ }\bibfield  {title} {\bibinfo {title} {{Exploiting classical
  nucleation theory for reverse self-assembly}},\ }\href
  {http://aip.scitation.org/doi/10.1063/1.3524307} {\bibfield  {journal}
  {\bibinfo  {journal} {J. Chem. Phys.}\ }\textbf {\bibinfo {volume} {133}},\
  \bibinfo {pages} {234108} (\bibinfo {year} {2010})}\BibitemShut {NoStop}%
\bibitem [{\citenamefont {Lin}\ \emph {et~al.}(1989)\citenamefont {Lin},
  \citenamefont {Lindsay}, \citenamefont {Weitz}, \citenamefont {Ball},
  \citenamefont {Klein},\ and\ \citenamefont {Meakin}}]{Lin1989}%
  \BibitemOpen
  \bibfield  {author} {\bibinfo {author} {\bibfnamefont {M.~Y.}\ \bibnamefont
  {Lin}}, \bibinfo {author} {\bibfnamefont {H.~M.}\ \bibnamefont {Lindsay}},
  \bibinfo {author} {\bibfnamefont {D.~A.}\ \bibnamefont {Weitz}}, \bibinfo
  {author} {\bibfnamefont {R.~C.}\ \bibnamefont {Ball}}, \bibinfo {author}
  {\bibfnamefont {R.}~\bibnamefont {Klein}},\ and\ \bibinfo {author}
  {\bibfnamefont {P.}~\bibnamefont {Meakin}},\ }\bibfield  {title} {\bibinfo
  {title} {{Universality in colloid aggregation}},\ }\href@noop {} {\bibfield
  {journal} {\bibinfo  {journal} {Nature}\ }\textbf {\bibinfo {volume} {339}},\
  \bibinfo {pages} {360} (\bibinfo {year} {1989})}\BibitemShut {NoStop}%
\bibitem [{\citenamefont {Sciortino}\ \emph {et~al.}(2004)\citenamefont
  {Sciortino}, \citenamefont {Mossa}, \citenamefont {Zaccarelli},\ and\
  \citenamefont {Tartaglia}}]{Sciortino2004}%
  \BibitemOpen
  \bibfield  {author} {\bibinfo {author} {\bibfnamefont {F.}~\bibnamefont
  {Sciortino}}, \bibinfo {author} {\bibfnamefont {S.}~\bibnamefont {Mossa}},
  \bibinfo {author} {\bibfnamefont {E.}~\bibnamefont {Zaccarelli}},\ and\
  \bibinfo {author} {\bibfnamefont {P.}~\bibnamefont {Tartaglia}},\ }\bibfield
  {title} {\bibinfo {title} {{Equilibrium cluster phases and low-density
  arrested disordered states: The role of short-range attraction and long-range
  repulsion}},\ }\href {https://link.aps.org/doi/10.1103/PhysRevLett.93.055701
  https://journals.aps.org/prl/abstract/10.1103/PhysRevLett.93.055701}
  {\bibfield  {journal} {\bibinfo  {journal} {Phys. Rev. Lett.}\ }\textbf
  {\bibinfo {volume} {93}},\ \bibinfo {pages} {55701} (\bibinfo {year}
  {2004})}\BibitemShut {NoStop}%
\bibitem [{\citenamefont {Charbonneau}\ and\ \citenamefont
  {Frenkel}(2007)}]{Charbonneau2007}%
  \BibitemOpen
  \bibfield  {author} {\bibinfo {author} {\bibfnamefont {P.}~\bibnamefont
  {Charbonneau}}\ and\ \bibinfo {author} {\bibfnamefont {D.}~\bibnamefont
  {Frenkel}},\ }\bibfield  {title} {\bibinfo {title} {Gas-solid coexistence of
  adhesive spheres},\ }\href@noop {} {\bibfield  {journal} {\bibinfo  {journal}
  {J. Chem. Phys.}\ }\textbf {\bibinfo {volume} {126}},\ \bibinfo {pages}
  {196101} (\bibinfo {year} {2007})}\BibitemShut {NoStop}%
\bibitem [{\citenamefont {Lu}\ \emph {et~al.}(2008)\citenamefont {Lu},
  \citenamefont {Zaccarelli}, \citenamefont {Ciulla}, \citenamefont
  {Schofield}, \citenamefont {Sciortino},\ and\ \citenamefont
  {Weitz}}]{Lu2008}%
  \BibitemOpen
  \bibfield  {author} {\bibinfo {author} {\bibfnamefont {P.~J.}\ \bibnamefont
  {Lu}}, \bibinfo {author} {\bibfnamefont {E.}~\bibnamefont {Zaccarelli}},
  \bibinfo {author} {\bibfnamefont {F.}~\bibnamefont {Ciulla}}, \bibinfo
  {author} {\bibfnamefont {A.~B.}\ \bibnamefont {Schofield}}, \bibinfo {author}
  {\bibfnamefont {F.}~\bibnamefont {Sciortino}},\ and\ \bibinfo {author}
  {\bibfnamefont {D.~A.}\ \bibnamefont {Weitz}},\ }\bibfield  {title} {\bibinfo
  {title} {{Gelation of particles with short-range attraction}},\ }\href@noop
  {} {\bibfield  {journal} {\bibinfo  {journal} {Nature}\ }\textbf {\bibinfo
  {volume} {453}},\ \bibinfo {pages} {499} (\bibinfo {year}
  {2008})}\BibitemShut {NoStop}%
\bibitem [{\citenamefont {Toledano}\ \emph {et~al.}(2009)\citenamefont
  {Toledano}, \citenamefont {Sciortino},\ and\ \citenamefont
  {Zaccarelli}}]{fernandeztoledano_colloidal_2009}%
  \BibitemOpen
  \bibfield  {author} {\bibinfo {author} {\bibfnamefont {J.~C.~F.}\
  \bibnamefont {Toledano}}, \bibinfo {author} {\bibfnamefont {F.}~\bibnamefont
  {Sciortino}},\ and\ \bibinfo {author} {\bibfnamefont {E.}~\bibnamefont
  {Zaccarelli}},\ }\bibfield  {title} {\bibinfo {title} {{Colloidal systems
  with competing interactions: From an arrested repulsive cluster phase to a
  gel}},\ }\href
  {http://pubs.rsc.org/en/Content/ArticleLanding/2009/SM/B818169A} {\bibfield
  {journal} {\bibinfo  {journal} {Soft Matter}\ }\textbf {\bibinfo {volume}
  {5}},\ \bibinfo {pages} {2390} (\bibinfo {year} {2009})}\BibitemShut
  {NoStop}%
\bibitem [{\citenamefont {Sanchez}\ and\ \citenamefont
  {Boening}(2014)}]{Sanchez2014}%
  \BibitemOpen
  \bibfield  {author} {\bibinfo {author} {\bibfnamefont {I.~C.}\ \bibnamefont
  {Sanchez}}\ and\ \bibinfo {author} {\bibfnamefont {K.~L.}\ \bibnamefont
  {Boening}},\ }\bibfield  {title} {\bibinfo {title} {{Universal thermodynamics
  at the liquid-vapor critical point}},\ }\href
  {http://pubs.acs.org/doi/10.1021/jp510096e} {\bibfield  {journal} {\bibinfo
  {journal} {J. Phys. Chem. B}\ }\textbf {\bibinfo {volume} {118}},\ \bibinfo
  {pages} {13704} (\bibinfo {year} {2014})}\BibitemShut {NoStop}%
\bibitem [{\citenamefont {Yan}\ \emph {et~al.}(2016)\citenamefont {Yan},
  \citenamefont {Han}, \citenamefont {Zhang}, \citenamefont {Xu}, \citenamefont
  {Luijten},\ and\ \citenamefont {Granick}}]{Yan2016}%
  \BibitemOpen
  \bibfield  {author} {\bibinfo {author} {\bibfnamefont {J.}~\bibnamefont
  {Yan}}, \bibinfo {author} {\bibfnamefont {M.}~\bibnamefont {Han}}, \bibinfo
  {author} {\bibfnamefont {J.}~\bibnamefont {Zhang}}, \bibinfo {author}
  {\bibfnamefont {C.}~\bibnamefont {Xu}}, \bibinfo {author} {\bibfnamefont
  {E.}~\bibnamefont {Luijten}},\ and\ \bibinfo {author} {\bibfnamefont
  {S.}~\bibnamefont {Granick}},\ }\bibfield  {title} {\bibinfo {title}
  {{Reconfiguring active particles by electrostatic imbalance}},\ }\href
  {http://www.nature.com/articles/nmat4696} {\bibfield  {journal} {\bibinfo
  {journal} {Nat. Mater.}\ }\textbf {\bibinfo {volume} {15}},\ \bibinfo {pages}
  {1095} (\bibinfo {year} {2016})}\BibitemShut {NoStop}%
\bibitem [{\citenamefont {Wang}\ \emph {et~al.}(2015)\citenamefont {Wang},
  \citenamefont {Duan}, \citenamefont {Ahmed}, \citenamefont {Sen},\ and\
  \citenamefont {Mallouk}}]{Wang2015}%
  \BibitemOpen
  \bibfield  {author} {\bibinfo {author} {\bibfnamefont {W.}~\bibnamefont
  {Wang}}, \bibinfo {author} {\bibfnamefont {W.}~\bibnamefont {Duan}}, \bibinfo
  {author} {\bibfnamefont {S.}~\bibnamefont {Ahmed}}, \bibinfo {author}
  {\bibfnamefont {A.}~\bibnamefont {Sen}},\ and\ \bibinfo {author}
  {\bibfnamefont {T.~E.}\ \bibnamefont {Mallouk}},\ }\bibfield  {title}
  {\bibinfo {title} {{From one to many: Dynamic assembly and collective
  behavior of self-propelled colloidal motors}},\ }\href
  {http://pubs.acs.org/doi/10.1021/acs.accounts.5b00025} {\bibfield  {journal}
  {\bibinfo  {journal} {Acc. Chem. Res.}\ }\textbf {\bibinfo {volume} {48}},\
  \bibinfo {pages} {1938} (\bibinfo {year} {2015})}\BibitemShut {NoStop}%
\bibitem [{\citenamefont {Gao}\ and\ \citenamefont {Wang}(2014)}]{Gao2014}%
  \BibitemOpen
  \bibfield  {author} {\bibinfo {author} {\bibfnamefont {W.}~\bibnamefont
  {Gao}}\ and\ \bibinfo {author} {\bibfnamefont {J.}~\bibnamefont {Wang}},\
  }\bibfield  {title} {\bibinfo {title} {{The Environmental Impact of
  Micro/Nanomachines: A Review}},\ }\href
  {http://pubs.acs.org/doi/10.1021/nn500077a} {\bibfield  {journal} {\bibinfo
  {journal} {ACS Nano}\ }\textbf {\bibinfo {volume} {8}},\ \bibinfo {pages}
  {3170} (\bibinfo {year} {2014})}\BibitemShut {NoStop}%
\bibitem [{\citenamefont {Brown}\ and\ \citenamefont {Poon}(2014)}]{Brown2014}%
  \BibitemOpen
  \bibfield  {author} {\bibinfo {author} {\bibfnamefont {A.}~\bibnamefont
  {Brown}}\ and\ \bibinfo {author} {\bibfnamefont {W.}~\bibnamefont {Poon}},\
  }\bibfield  {title} {\bibinfo {title} {{Ionic effects in self-propelled
  Pt-coated Janus swimmers}},\ }\href@noop {} {\bibfield  {journal} {\bibinfo
  {journal} {Soft Matter}\ }\textbf {\bibinfo {volume} {10}},\ \bibinfo {pages}
  {4016} (\bibinfo {year} {2014})}\BibitemShut {NoStop}%
\bibitem [{\citenamefont {Palacci}\ \emph {et~al.}(2013)\citenamefont
  {Palacci}, \citenamefont {Sacanna}, \citenamefont {Vatchinsky}, \citenamefont
  {Chaikin},\ and\ \citenamefont {Pine}}]{Palacci2013a}%
  \BibitemOpen
  \bibfield  {author} {\bibinfo {author} {\bibfnamefont {J.}~\bibnamefont
  {Palacci}}, \bibinfo {author} {\bibfnamefont {S.}~\bibnamefont {Sacanna}},
  \bibinfo {author} {\bibfnamefont {A.}~\bibnamefont {Vatchinsky}}, \bibinfo
  {author} {\bibfnamefont {P.~M.}\ \bibnamefont {Chaikin}},\ and\ \bibinfo
  {author} {\bibfnamefont {D.~J.}\ \bibnamefont {Pine}},\ }\bibfield  {title}
  {\bibinfo {title} {{Photoactivated colloidal dockers for cargo
  transportation}},\ }\href@noop {} {\bibfield  {journal} {\bibinfo  {journal}
  {J. Am. Chem. Soc.}\ }\textbf {\bibinfo {volume} {135}},\ \bibinfo {pages}
  {15978} (\bibinfo {year} {2013})}\BibitemShut {NoStop}%
\bibitem [{\citenamefont {Wang}\ \emph {et~al.}(2013)\citenamefont {Wang},
  \citenamefont {Duan}, \citenamefont {Ahmed}, \citenamefont {Mallouk},\ and\
  \citenamefont {Sen}}]{wang_small_2013}%
  \BibitemOpen
  \bibfield  {author} {\bibinfo {author} {\bibfnamefont {W.}~\bibnamefont
  {Wang}}, \bibinfo {author} {\bibfnamefont {W.}~\bibnamefont {Duan}}, \bibinfo
  {author} {\bibfnamefont {S.}~\bibnamefont {Ahmed}}, \bibinfo {author}
  {\bibfnamefont {T.~E.}\ \bibnamefont {Mallouk}},\ and\ \bibinfo {author}
  {\bibfnamefont {A.}~\bibnamefont {Sen}},\ }\bibfield  {title} {\bibinfo
  {title} {{Small power: Autonomous nano- and micromotors propelled by
  self-generated gradients}},\ }\href
  {http://www.sciencedirect.com/science/article/pii/S1748013213000947}
  {\bibfield  {journal} {\bibinfo  {journal} {Nano Today}\ }\textbf {\bibinfo
  {volume} {8}},\ \bibinfo {pages} {531} (\bibinfo {year} {2013})}\BibitemShut
  {NoStop}%
\bibitem [{\citenamefont {Dey}\ and\ \citenamefont
  {Sen}(2017)}]{dey_chemically_2017}%
  \BibitemOpen
  \bibfield  {author} {\bibinfo {author} {\bibfnamefont {K.~K.}\ \bibnamefont
  {Dey}}\ and\ \bibinfo {author} {\bibfnamefont {A.}~\bibnamefont {Sen}},\
  }\bibfield  {title} {\bibinfo {title} {{Chemically Propelled Molecules and
  Machines}},\ }\href {http://dx.doi.org/10.1021/jacs.7b02347} {\bibfield
  {journal} {\bibinfo  {journal} {J. Am. Chem. Soc.}\ }\textbf {\bibinfo
  {volume} {139}},\ \bibinfo {pages} {7666} (\bibinfo {year}
  {2017})}\BibitemShut {NoStop}%
\bibitem [{\citenamefont {Ramaswamy}(2010)}]{Ramaswamy2010}%
  \BibitemOpen
  \bibfield  {author} {\bibinfo {author} {\bibfnamefont {S.}~\bibnamefont
  {Ramaswamy}},\ }\bibfield  {title} {\bibinfo {title} {{The Mechanics and
  Statistics of Active Matter}},\ }\href
  {http://www.annualreviews.org/doi/10.1146/annurev-conmatphys-070909-104101
  http://arxiv.org/abs/1004.1933{\%}0Ahttp://dx.doi.org/10.1146/annurev-conmatphys-070909-104101}
  {\bibfield  {journal} {\bibinfo  {journal} {Annu. Rev. Condens. Matter
  Phys.}\ }\textbf {\bibinfo {volume} {1}},\ \bibinfo {pages} {323} (\bibinfo
  {year} {2010})}\BibitemShut {NoStop}%
\bibitem [{\citenamefont {Bechinger}\ \emph {et~al.}(2016)\citenamefont
  {Bechinger}, \citenamefont {{Di Leonardo}}, \citenamefont {L{\"{o}}wen},
  \citenamefont {Reichhardt}, \citenamefont {Volpe},\ and\ \citenamefont
  {Volpe}}]{Bechinger2016}%
  \BibitemOpen
  \bibfield  {author} {\bibinfo {author} {\bibfnamefont {C.}~\bibnamefont
  {Bechinger}}, \bibinfo {author} {\bibfnamefont {R.}~\bibnamefont {{Di
  Leonardo}}}, \bibinfo {author} {\bibfnamefont {H.}~\bibnamefont
  {L{\"{o}}wen}}, \bibinfo {author} {\bibfnamefont {C.}~\bibnamefont
  {Reichhardt}}, \bibinfo {author} {\bibfnamefont {G.}~\bibnamefont {Volpe}},\
  and\ \bibinfo {author} {\bibfnamefont {G.}~\bibnamefont {Volpe}},\ }\bibfield
   {title} {\bibinfo {title} {{Active particles in complex and crowded
  environments}},\ }\href {http://link.aps.org/doi/10.1103/RevModPhys.88.045006
  http://journals.aps.org/rmp/abstract/10.1103/RevModPhys.88.045006} {\bibfield
   {journal} {\bibinfo  {journal} {Rev. Mod. Phys.}\ }\textbf {\bibinfo
  {volume} {88}},\ \bibinfo {pages} {45006} (\bibinfo {year}
  {2016})}\BibitemShut {NoStop}%
\bibitem [{\citenamefont {{Di Leonardo}}(2016)}]{DiLeonardo2016}%
  \BibitemOpen
  \bibfield  {author} {\bibinfo {author} {\bibfnamefont {R.}~\bibnamefont {{Di
  Leonardo}}},\ }\bibfield  {title} {\bibinfo {title} {{Active colloids:
  Controlled collective motions}},\ }\href
  {http://www.nature.com/doifinder/10.1038/nmat4761} {\bibfield  {journal}
  {\bibinfo  {journal} {Nat. Mater.}\ }\textbf {\bibinfo {volume} {15}},\
  \bibinfo {pages} {1057} (\bibinfo {year} {2016})}\BibitemShut {NoStop}%
\bibitem [{\citenamefont {Patteson}\ \emph {et~al.}(2016)\citenamefont
  {Patteson}, \citenamefont {Gopinath},\ and\ \citenamefont
  {Arratia}}]{Patteson2016}%
  \BibitemOpen
  \bibfield  {author} {\bibinfo {author} {\bibfnamefont {A.~E.}\ \bibnamefont
  {Patteson}}, \bibinfo {author} {\bibfnamefont {A.}~\bibnamefont {Gopinath}},\
  and\ \bibinfo {author} {\bibfnamefont {P.~E.}\ \bibnamefont {Arratia}},\
  }\bibfield  {title} {\bibinfo {title} {{Active colloids in complex fluids}},\
  }\href {http://www.sciencedirect.com/science/article/pii/S1359029416000030}
  {\bibfield  {journal} {\bibinfo  {journal} {Curr. Opin. Colloid Interface
  Sci.}\ }\textbf {\bibinfo {volume} {21}},\ \bibinfo {pages} {86} (\bibinfo
  {year} {2016})}\BibitemShut {NoStop}%
\bibitem [{\citenamefont {Z{\"{o}}ttl}\ and\ \citenamefont
  {Stark}(2016)}]{zottl_emergent_2016}%
  \BibitemOpen
  \bibfield  {author} {\bibinfo {author} {\bibfnamefont {A.}~\bibnamefont
  {Z{\"{o}}ttl}}\ and\ \bibinfo {author} {\bibfnamefont {H.}~\bibnamefont
  {Stark}},\ }\bibfield  {title} {\bibinfo {title} {{Emergent behavior in
  active colloids}},\ }\href {http://stacks.iop.org/0953-8984/28/i=25/a=253001
  http://stacks.iop.org/0953-8984/28/i=25/a=253001?key=crossref.4f31c1a4927a885005054f0c7ca05d87}
  {\bibfield  {journal} {\bibinfo  {journal} {J. Phys. Condens. Matter}\
  }\textbf {\bibinfo {volume} {28}},\ \bibinfo {pages} {253001} (\bibinfo
  {year} {2016})}\BibitemShut {NoStop}%
\bibitem [{\citenamefont {Cates}\ and\ \citenamefont
  {Tailleur}(2015)}]{Cates2015}%
  \BibitemOpen
  \bibfield  {author} {\bibinfo {author} {\bibfnamefont {M.~E.}\ \bibnamefont
  {Cates}}\ and\ \bibinfo {author} {\bibfnamefont {J.}~\bibnamefont
  {Tailleur}},\ }\bibfield  {title} {\bibinfo {title} {{Motility-Induced Phase
  Separation}},\ }\href@noop {} {\bibfield  {journal} {\bibinfo  {journal}
  {Annu. Rev. Condens. Matter Phys.}\ }\textbf {\bibinfo {volume} {6}},\
  \bibinfo {pages} {219} (\bibinfo {year} {2015})}\BibitemShut {NoStop}%
\bibitem [{\citenamefont {Dauchot}\ and\ \citenamefont
  {L{\"{o}}wen}(2019)}]{Dauchot2019}%
  \BibitemOpen
  \bibfield  {author} {\bibinfo {author} {\bibfnamefont {O.}~\bibnamefont
  {Dauchot}}\ and\ \bibinfo {author} {\bibfnamefont {H.}~\bibnamefont
  {L{\"{o}}wen}},\ }\bibfield  {title} {\bibinfo {title} {{Chemical Physics of
  Active Matter}},\ }\href@noop {} {\bibfield  {journal} {\bibinfo  {journal}
  {J. Chem. Phys.}\ }\textbf {\bibinfo {volume} {151}},\ \bibinfo {pages}
  {114901} (\bibinfo {year} {2019})}\BibitemShut {NoStop}%
\bibitem [{\citenamefont {Speck}(2020)}]{Speck2020}%
  \BibitemOpen
  \bibfield  {author} {\bibinfo {author} {\bibfnamefont {T.}~\bibnamefont
  {Speck}},\ }\bibfield  {title} {\bibinfo {title} {Collective forces in scalar
  active matter},\ }\href {https://doi.org/10.1039/D0SM00176G} {\bibfield
  {journal} {\bibinfo  {journal} {Soft Matter}\ }\textbf {\bibinfo {volume}
  {16}},\ \bibinfo {pages} {2652} (\bibinfo {year} {2020})}\BibitemShut
  {NoStop}%
\bibitem [{\citenamefont {{Marini Bettolo Marconi}}\ and\ \citenamefont
  {Maggi}(2015)}]{MariniBettoloMarconi2015}%
  \BibitemOpen
  \bibfield  {author} {\bibinfo {author} {\bibfnamefont {U.}~\bibnamefont
  {{Marini Bettolo Marconi}}}\ and\ \bibinfo {author} {\bibfnamefont
  {C.}~\bibnamefont {Maggi}},\ }\bibfield  {title} {\bibinfo {title} {{Towards
  a statistical mechanical theory of active fluids}},\ }\href
  {http://xlink.rsc.org/?DOI=C5SM01718A} {\bibfield  {journal} {\bibinfo
  {journal} {Soft Matter}\ }\textbf {\bibinfo {volume} {11}},\ \bibinfo {pages}
  {8768} (\bibinfo {year} {2015})}\BibitemShut {NoStop}%
\bibitem [{\citenamefont {Mallory}\ \emph
  {et~al.}(2017{\natexlab{a}})\citenamefont {Mallory}, \citenamefont
  {Valeriani},\ and\ \citenamefont {Cacciuto}}]{Mallory2017a}%
  \BibitemOpen
  \bibfield  {author} {\bibinfo {author} {\bibfnamefont {S.~A.}\ \bibnamefont
  {Mallory}}, \bibinfo {author} {\bibfnamefont {C.}~\bibnamefont {Valeriani}},\
  and\ \bibinfo {author} {\bibfnamefont {A.}~\bibnamefont {Cacciuto}},\
  }\bibfield  {title} {\bibinfo {title} {{An Active Approach to Colloidal
  Self-Assembly}},\ }\href
  {http://www.annualreviews.org/doi/10.1146/annurev-physchem-050317-021237}
  {\bibfield  {journal} {\bibinfo  {journal} {Annu. Rev. Phys. Chem.}\ }\textbf
  {\bibinfo {volume} {69}},\ \bibinfo {pages} {59} (\bibinfo {year}
  {2017}{\natexlab{a}})}\BibitemShut {NoStop}%
\bibitem [{\citenamefont {Das}\ \emph {et~al.}(2019)\citenamefont {Das},
  \citenamefont {{Lee Bowers}}, \citenamefont {Bakker},\ and\ \citenamefont
  {Cacciuto}}]{Das2019}%
  \BibitemOpen
  \bibfield  {author} {\bibinfo {author} {\bibfnamefont {S.}~\bibnamefont
  {Das}}, \bibinfo {author} {\bibfnamefont {M.}~\bibnamefont {{Lee Bowers}}},
  \bibinfo {author} {\bibfnamefont {C.}~\bibnamefont {Bakker}},\ and\ \bibinfo
  {author} {\bibfnamefont {A.}~\bibnamefont {Cacciuto}},\ }\bibfield  {title}
  {\bibinfo {title} {{Active sculpting of colloidal crystals}},\ }\href
  {http://aip.scitation.org/doi/10.1063/1.5082949} {\bibfield  {journal}
  {\bibinfo  {journal} {J. Chem. Phys.}\ }\textbf {\bibinfo {volume} {150}},\
  \bibinfo {pages} {134505} (\bibinfo {year} {2019})}\BibitemShut {NoStop}%
\bibitem [{\citenamefont {Shan}\ \emph {et~al.}(2019)\citenamefont {Shan},
  \citenamefont {Zhang}, \citenamefont {Tian},\ and\ \citenamefont
  {Chen}}]{Shan2019}%
  \BibitemOpen
  \bibfield  {author} {\bibinfo {author} {\bibfnamefont {W.~J.}\ \bibnamefont
  {Shan}}, \bibinfo {author} {\bibfnamefont {F.}~\bibnamefont {Zhang}},
  \bibinfo {author} {\bibfnamefont {W.~D.}\ \bibnamefont {Tian}},\ and\
  \bibinfo {author} {\bibfnamefont {K.}~\bibnamefont {Chen}},\ }\bibfield
  {title} {\bibinfo {title} {{Assembly structures and dynamics of active
  colloidal cells}},\ }\href@noop {} {\bibfield  {journal} {\bibinfo  {journal}
  {Soft Matter}\ }\textbf {\bibinfo {volume} {15}},\ \bibinfo {pages} {4761}
  (\bibinfo {year} {2019})}\BibitemShut {NoStop}%
\bibitem [{\citenamefont {Prymidis}\ \emph {et~al.}(2015)\citenamefont
  {Prymidis}, \citenamefont {Sielcken},\ and\ \citenamefont
  {Filion}}]{Prymidis2015}%
  \BibitemOpen
  \bibfield  {author} {\bibinfo {author} {\bibfnamefont {V.}~\bibnamefont
  {Prymidis}}, \bibinfo {author} {\bibfnamefont {H.}~\bibnamefont {Sielcken}},\
  and\ \bibinfo {author} {\bibfnamefont {L.}~\bibnamefont {Filion}},\
  }\bibfield  {title} {\bibinfo {title} {{Self-assembly of active attractive
  spheres}},\ }\href
  {http://pubs.rsc.org/en/content/articlelanding/2015/sm/c5sm00127g
  http://pubs.rsc.org/en/content/articlepdf/2015/sm/c5sm00127g
  http://pubs.rsc.org/en/Content/ArticlePDF/2015/SM/C5SM00127G
  http://pubs.rsc.org/en/Content/ArticleLanding/2015/SM/C5SM00127G{\#}!di}
  {\bibfield  {journal} {\bibinfo  {journal} {Soft Matter}\ }\textbf {\bibinfo
  {volume} {11}},\ \bibinfo {pages} {4158} (\bibinfo {year}
  {2015})}\BibitemShut {NoStop}%
\bibitem [{\citenamefont {Solovev}\ \emph {et~al.}(2012)\citenamefont
  {Solovev}, \citenamefont {Xi}, \citenamefont {Gracias}, \citenamefont
  {Harazim}, \citenamefont {Deneke}, \citenamefont {Sanchez},\ and\
  \citenamefont {Schmidt}}]{solovev_self-propelled_2012}%
  \BibitemOpen
  \bibfield  {author} {\bibinfo {author} {\bibfnamefont {A.~A.}\ \bibnamefont
  {Solovev}}, \bibinfo {author} {\bibfnamefont {W.}~\bibnamefont {Xi}},
  \bibinfo {author} {\bibfnamefont {D.~H.}\ \bibnamefont {Gracias}}, \bibinfo
  {author} {\bibfnamefont {S.~M.}\ \bibnamefont {Harazim}}, \bibinfo {author}
  {\bibfnamefont {C.}~\bibnamefont {Deneke}}, \bibinfo {author} {\bibfnamefont
  {S.}~\bibnamefont {Sanchez}},\ and\ \bibinfo {author} {\bibfnamefont {O.~G.}\
  \bibnamefont {Schmidt}},\ }\bibfield  {title} {\bibinfo {title}
  {{Self-propelled nanotools}},\ }\href {http://dx.doi.org/10.1021/nn204762w}
  {\bibfield  {journal} {\bibinfo  {journal} {ACS Nano}\ }\textbf {\bibinfo
  {volume} {6}},\ \bibinfo {pages} {1751} (\bibinfo {year} {2012})}\BibitemShut
  {NoStop}%
\bibitem [{\citenamefont {Wang}\ \emph
  {et~al.}(2019{\natexlab{a}})\citenamefont {Wang}, \citenamefont {Guo},
  \citenamefont {Tian},\ and\ \citenamefont {Chen}}]{Wang2019a}%
  \BibitemOpen
  \bibfield  {author} {\bibinfo {author} {\bibfnamefont {C.}~\bibnamefont
  {Wang}}, \bibinfo {author} {\bibfnamefont {Y.~K.}\ \bibnamefont {Guo}},
  \bibinfo {author} {\bibfnamefont {W.~D.}\ \bibnamefont {Tian}},\ and\
  \bibinfo {author} {\bibfnamefont {K.}~\bibnamefont {Chen}},\ }\bibfield
  {title} {\bibinfo {title} {{Shape transformation and manipulation of a
  vesicle by active particles}},\ }\href
  {http://aip.scitation.org/doi/10.1063/1.5078694} {\bibfield  {journal}
  {\bibinfo  {journal} {J. Chem. Phys.}\ }\textbf {\bibinfo {volume} {150}},\
  \bibinfo {pages} {044907} (\bibinfo {year} {2019}{\natexlab{a}})}\BibitemShut
  {NoStop}%
\bibitem [{\citenamefont {Angelani}(2019)}]{Angelani2019}%
  \BibitemOpen
  \bibfield  {author} {\bibinfo {author} {\bibfnamefont {L.}~\bibnamefont
  {Angelani}},\ }\bibfield  {title} {\bibinfo {title} {Spontaneous assembly of
  colloidal vesicles driven by active swimmers},\ }\href@noop {} {\bibfield
  {journal} {\bibinfo  {journal} {J. Phys. Condens. Matter}\ }\textbf {\bibinfo
  {volume} {31}},\ \bibinfo {pages} {075101} (\bibinfo {year}
  {2019})}\BibitemShut {NoStop}%
\bibitem [{\citenamefont {Gr{\"{u}}nwald}\ \emph {et~al.}(2016)\citenamefont
  {Gr{\"{u}}nwald}, \citenamefont {Tricard}, \citenamefont {Whitesides},\ and\
  \citenamefont {Geissler}}]{Grunwald2016}%
  \BibitemOpen
  \bibfield  {author} {\bibinfo {author} {\bibfnamefont {M.}~\bibnamefont
  {Gr{\"{u}}nwald}}, \bibinfo {author} {\bibfnamefont {S.}~\bibnamefont
  {Tricard}}, \bibinfo {author} {\bibfnamefont {G.~M.}\ \bibnamefont
  {Whitesides}},\ and\ \bibinfo {author} {\bibfnamefont {P.~L.}\ \bibnamefont
  {Geissler}},\ }\bibfield  {title} {\bibinfo {title} {{Exploiting
  non-equilibrium phase separation for self-assembly}},\ }\href@noop {}
  {\bibfield  {journal} {\bibinfo  {journal} {Soft Matter}\ }\textbf {\bibinfo
  {volume} {12}},\ \bibinfo {pages} {1517} (\bibinfo {year}
  {2016})}\BibitemShut {NoStop}%
\bibitem [{\citenamefont {Maggi}\ \emph {et~al.}(2016)\citenamefont {Maggi},
  \citenamefont {Simmchen}, \citenamefont {Saglimbeni}, \citenamefont {Katuri},
  \citenamefont {Dipalo}, \citenamefont {{De Angelis}}, \citenamefont
  {Sanchez},\ and\ \citenamefont {{Di Leonardo}}}]{maggi_self-assembly_2016}%
  \BibitemOpen
  \bibfield  {author} {\bibinfo {author} {\bibfnamefont {C.}~\bibnamefont
  {Maggi}}, \bibinfo {author} {\bibfnamefont {J.}~\bibnamefont {Simmchen}},
  \bibinfo {author} {\bibfnamefont {F.}~\bibnamefont {Saglimbeni}}, \bibinfo
  {author} {\bibfnamefont {J.}~\bibnamefont {Katuri}}, \bibinfo {author}
  {\bibfnamefont {M.}~\bibnamefont {Dipalo}}, \bibinfo {author} {\bibfnamefont
  {F.}~\bibnamefont {{De Angelis}}}, \bibinfo {author} {\bibfnamefont
  {S.}~\bibnamefont {Sanchez}},\ and\ \bibinfo {author} {\bibfnamefont
  {R.}~\bibnamefont {{Di Leonardo}}},\ }\bibfield  {title} {\bibinfo {title}
  {{Self-Assembly of Micromachining Systems Powered by Janus Micromotors}},\
  }\href {http://onlinelibrary.wiley.com/doi/10.1002/smll.201502391/abstract}
  {\bibfield  {journal} {\bibinfo  {journal} {Small}\ }\textbf {\bibinfo
  {volume} {12}},\ \bibinfo {pages} {446} (\bibinfo {year} {2016})}\BibitemShut
  {NoStop}%
\bibitem [{\citenamefont {Wang}\ and\ \citenamefont
  {Simmchen}(2019)}]{wang2019interactions}%
  \BibitemOpen
  \bibfield  {author} {\bibinfo {author} {\bibfnamefont {L.}~\bibnamefont
  {Wang}}\ and\ \bibinfo {author} {\bibfnamefont {J.}~\bibnamefont
  {Simmchen}},\ }\bibfield  {title} {\bibinfo {title} {Interactions of active
  colloids with passive tracers},\ }\href@noop {} {\bibfield  {journal}
  {\bibinfo  {journal} {Condensed Matter}\ }\textbf {\bibinfo {volume} {4}},\
  \bibinfo {pages} {78} (\bibinfo {year} {2019})}\BibitemShut {NoStop}%
\bibitem [{\citenamefont {Mallory}\ and\ \citenamefont
  {Cacciuto}(2019)}]{Mallory2019}%
  \BibitemOpen
  \bibfield  {author} {\bibinfo {author} {\bibfnamefont {S.~A.}\ \bibnamefont
  {Mallory}}\ and\ \bibinfo {author} {\bibfnamefont {A.}~\bibnamefont
  {Cacciuto}},\ }\bibfield  {title} {\bibinfo {title} {Activity-enhanced
  self-assembly of a colloidal kagome lattice},\ }\href@noop {} {\bibfield
  {journal} {\bibinfo  {journal} {J. Am. Chem. Soc.}\ }\textbf {\bibinfo
  {volume} {141}},\ \bibinfo {pages} {2500} (\bibinfo {year}
  {2019})}\BibitemShut {NoStop}%
\bibitem [{\citenamefont {Wensink}\ \emph {et~al.}(2014)\citenamefont
  {Wensink}, \citenamefont {Kantsler}, \citenamefont {Goldstein},\ and\
  \citenamefont {Dunkel}}]{Wensink2014}%
  \BibitemOpen
  \bibfield  {author} {\bibinfo {author} {\bibfnamefont {H.~H.}\ \bibnamefont
  {Wensink}}, \bibinfo {author} {\bibfnamefont {V.}~\bibnamefont {Kantsler}},
  \bibinfo {author} {\bibfnamefont {R.~E.}\ \bibnamefont {Goldstein}},\ and\
  \bibinfo {author} {\bibfnamefont {J.}~\bibnamefont {Dunkel}},\ }\bibfield
  {title} {\bibinfo {title} {{Controlling active self-assembly through broken
  particle-shape symmetry}},\ }\href@noop {} {\bibfield  {journal} {\bibinfo
  {journal} {Phys. Rev. E - Stat. Nonlinear, Soft Matter Phys.}\ }\textbf
  {\bibinfo {volume} {89}},\ \bibinfo {pages} {10302} (\bibinfo {year}
  {2014})}\BibitemShut {NoStop}%
\bibitem [{\citenamefont {Aldana}\ \emph {et~al.}(2020)\citenamefont {Aldana},
  \citenamefont {Fuentes-Cabrera},\ and\ \citenamefont {Zumaya}}]{Aldana2020}%
  \BibitemOpen
  \bibfield  {author} {\bibinfo {author} {\bibfnamefont {M.}~\bibnamefont
  {Aldana}}, \bibinfo {author} {\bibfnamefont {M.}~\bibnamefont
  {Fuentes-Cabrera}},\ and\ \bibinfo {author} {\bibfnamefont {M.}~\bibnamefont
  {Zumaya}},\ }\bibfield  {title} {\bibinfo {title} {{Self-Propulsion Enhances
  Polymerization}},\ }\href {https://www.mdpi.com/1099-4300/22/2/251}
  {\bibfield  {journal} {\bibinfo  {journal} {Entropy}\ }\textbf {\bibinfo
  {volume} {22}},\ \bibinfo {pages} {251} (\bibinfo {year} {2020})}\BibitemShut
  {NoStop}%
\bibitem [{\citenamefont {Mallory}\ \emph
  {et~al.}(2017{\natexlab{b}})\citenamefont {Mallory}, \citenamefont {Alarcon},
  \citenamefont {Cacciuto},\ and\ \citenamefont {Valeriani}}]{Mallory2017}%
  \BibitemOpen
  \bibfield  {author} {\bibinfo {author} {\bibfnamefont {S.}~\bibnamefont
  {Mallory}}, \bibinfo {author} {\bibfnamefont {F.}~\bibnamefont {Alarcon}},
  \bibinfo {author} {\bibfnamefont {A.}~\bibnamefont {Cacciuto}},\ and\
  \bibinfo {author} {\bibfnamefont {C.}~\bibnamefont {Valeriani}},\ }\bibfield
  {title} {\bibinfo {title} {Self-assembly of active amphiphilic janus
  particles},\ }\href@noop {} {\bibfield  {journal} {\bibinfo  {journal} {New
  J. Phys.}\ }\textbf {\bibinfo {volume} {19}},\ \bibinfo {pages} {125014}
  (\bibinfo {year} {2017}{\natexlab{b}})}\BibitemShut {NoStop}%
\bibitem [{\citenamefont {Mallory}\ and\ \citenamefont
  {Cacciuto}(2016)}]{Mallory2016}%
  \BibitemOpen
  \bibfield  {author} {\bibinfo {author} {\bibfnamefont {S.~A.}\ \bibnamefont
  {Mallory}}\ and\ \bibinfo {author} {\bibfnamefont {A.}~\bibnamefont
  {Cacciuto}},\ }\bibfield  {title} {\bibinfo {title} {Activity-assisted
  self-assembly of colloidal particles},\ }\href
  {https://doi.org/10.1103/PhysRevE.94.022607} {\bibfield  {journal} {\bibinfo
  {journal} {Phys. Rev. E}\ }\textbf {\bibinfo {volume} {94}},\ \bibinfo
  {pages} {022607} (\bibinfo {year} {2016})}\BibitemShut {NoStop}%
\bibitem [{\citenamefont {Hess}(2006)}]{Hess2006}%
  \BibitemOpen
  \bibfield  {author} {\bibinfo {author} {\bibfnamefont {H.}~\bibnamefont
  {Hess}},\ }\bibfield  {title} {\bibinfo {title} {{Self-assembly driven by
  molecular motors}},\ }\href {http://xlink.rsc.org/?DOI=b518281f} {\bibfield
  {journal} {\bibinfo  {journal} {Soft Matter}\ }\textbf {\bibinfo {volume}
  {2}},\ \bibinfo {pages} {669} (\bibinfo {year} {2006})}\BibitemShut {NoStop}%
\bibitem [{\citenamefont {{Davies Wykes}}\ \emph {et~al.}(2016)\citenamefont
  {{Davies Wykes}}, \citenamefont {Palacci}, \citenamefont {Adachi},
  \citenamefont {Ristroph}, \citenamefont {Zhong}, \citenamefont {Ward},
  \citenamefont {Zhang},\ and\ \citenamefont {Shelley}}]{wykes_dynamic_2016}%
  \BibitemOpen
  \bibfield  {author} {\bibinfo {author} {\bibfnamefont {M.~S.}\ \bibnamefont
  {{Davies Wykes}}}, \bibinfo {author} {\bibfnamefont {J.}~\bibnamefont
  {Palacci}}, \bibinfo {author} {\bibfnamefont {T.}~\bibnamefont {Adachi}},
  \bibinfo {author} {\bibfnamefont {L.}~\bibnamefont {Ristroph}}, \bibinfo
  {author} {\bibfnamefont {X.}~\bibnamefont {Zhong}}, \bibinfo {author}
  {\bibfnamefont {M.~D.}\ \bibnamefont {Ward}}, \bibinfo {author}
  {\bibfnamefont {J.}~\bibnamefont {Zhang}},\ and\ \bibinfo {author}
  {\bibfnamefont {M.~J.}\ \bibnamefont {Shelley}},\ }\bibfield  {title}
  {\bibinfo {title} {{Dynamic self-assembly of microscale rotors and
  swimmers}},\ }\href
  {http://pubs.rsc.org/en/content/articlelanding/2016/sm/c5sm03127c} {\bibfield
   {journal} {\bibinfo  {journal} {Soft Matter}\ }\textbf {\bibinfo {volume}
  {12}},\ \bibinfo {pages} {4584} (\bibinfo {year} {2016})}\BibitemShut
  {NoStop}%
\bibitem [{\citenamefont {Du}\ \emph {et~al.}(2019)\citenamefont {Du},
  \citenamefont {Jiang},\ and\ \citenamefont {Hou}}]{Du2019}%
  \BibitemOpen
  \bibfield  {author} {\bibinfo {author} {\bibfnamefont {Y.}~\bibnamefont
  {Du}}, \bibinfo {author} {\bibfnamefont {H.}~\bibnamefont {Jiang}},\ and\
  \bibinfo {author} {\bibfnamefont {Z.}~\bibnamefont {Hou}},\ }\bibfield
  {title} {\bibinfo {title} {{Self-assembly of active core corona particles
  into highly ordered and self-healing structures}},\ }\href
  {http://aip.scitation.org/doi/10.1063/1.5121802} {\bibfield  {journal}
  {\bibinfo  {journal} {J. Chem. Phys.}\ }\textbf {\bibinfo {volume} {151}},\
  \bibinfo {pages} {154904} (\bibinfo {year} {2019})}\BibitemShut {NoStop}%
\bibitem [{\citenamefont {Ni}\ \emph {et~al.}(2013)\citenamefont {Ni},
  \citenamefont {Stuart},\ and\ \citenamefont {Dijkstra}}]{ni2013pushing}%
  \BibitemOpen
  \bibfield  {author} {\bibinfo {author} {\bibfnamefont {R.}~\bibnamefont
  {Ni}}, \bibinfo {author} {\bibfnamefont {M.~A.~C.}\ \bibnamefont {Stuart}},\
  and\ \bibinfo {author} {\bibfnamefont {M.}~\bibnamefont {Dijkstra}},\
  }\bibfield  {title} {\bibinfo {title} {Pushing the glass transition towards
  random close packing using self-propelled hard spheres},\ }\href
  {https://doi.org/10.1038/ncomms3704} {\bibfield  {journal} {\bibinfo
  {journal} {Nature Communications}\ }\textbf {\bibinfo {volume} {4}},\
  \bibinfo {pages} {2704} (\bibinfo {year} {2013})}\BibitemShut {NoStop}%
\bibitem [{\citenamefont {Wang}\ \emph
  {et~al.}(2019{\natexlab{b}})\citenamefont {Wang}, \citenamefont {Wang},
  \citenamefont {Li}, \citenamefont {Cheung}, \citenamefont {Tian},
  \citenamefont {Kim}, \citenamefont {Yi}, \citenamefont {Ducrot},\ and\
  \citenamefont {Wang}}]{Wang2019}%
  \BibitemOpen
  \bibfield  {author} {\bibinfo {author} {\bibfnamefont {Z.}~\bibnamefont
  {Wang}}, \bibinfo {author} {\bibfnamefont {Z.}~\bibnamefont {Wang}}, \bibinfo
  {author} {\bibfnamefont {J.}~\bibnamefont {Li}}, \bibinfo {author}
  {\bibfnamefont {S.~T.~H.}\ \bibnamefont {Cheung}}, \bibinfo {author}
  {\bibfnamefont {C.}~\bibnamefont {Tian}}, \bibinfo {author} {\bibfnamefont
  {S.~H.}\ \bibnamefont {Kim}}, \bibinfo {author} {\bibfnamefont {G.~R.}\
  \bibnamefont {Yi}}, \bibinfo {author} {\bibfnamefont {E.}~\bibnamefont
  {Ducrot}},\ and\ \bibinfo {author} {\bibfnamefont {Y.}~\bibnamefont {Wang}},\
  }\bibfield  {title} {\bibinfo {title} {{Active Patchy Colloids with
  Shape-Tunable Dynamics}},\ }\href@noop {} {\bibfield  {journal} {\bibinfo
  {journal} {J. Am. Chem. Soc.}\ }\textbf {\bibinfo {volume} {141}},\ \bibinfo
  {pages} {14853} (\bibinfo {year} {2019}{\natexlab{b}})}\BibitemShut {NoStop}%
\bibitem [{\citenamefont {Gao}\ \emph {et~al.}(2013)\citenamefont {Gao},
  \citenamefont {Pei}, \citenamefont {Feng}, \citenamefont {Hennessy},\ and\
  \citenamefont {Wang}}]{Gao2013a}%
  \BibitemOpen
  \bibfield  {author} {\bibinfo {author} {\bibfnamefont {W.}~\bibnamefont
  {Gao}}, \bibinfo {author} {\bibfnamefont {A.}~\bibnamefont {Pei}}, \bibinfo
  {author} {\bibfnamefont {X.}~\bibnamefont {Feng}}, \bibinfo {author}
  {\bibfnamefont {C.}~\bibnamefont {Hennessy}},\ and\ \bibinfo {author}
  {\bibfnamefont {J.}~\bibnamefont {Wang}},\ }\bibfield  {title} {\bibinfo
  {title} {{Organized Self-Assembly of Janus Micromotors with Hydrophobic
  Hemispheres}},\ }\href {http://pubs.acs.org/doi/10.1021/ja311455k} {\bibfield
   {journal} {\bibinfo  {journal} {J. Am. Chem. Soc.}\ }\textbf {\bibinfo
  {volume} {135}},\ \bibinfo {pages} {998} (\bibinfo {year}
  {2013})}\BibitemShut {NoStop}%
\bibitem [{\citenamefont {Dietrich}\ \emph {et~al.}(2018)\citenamefont
  {Dietrich}, \citenamefont {Volpe}, \citenamefont {Sulaiman}, \citenamefont
  {Renggli}, \citenamefont {Buttinoni},\ and\ \citenamefont
  {Isa}}]{PhysRevLett.120.268004}%
  \BibitemOpen
  \bibfield  {author} {\bibinfo {author} {\bibfnamefont {K.}~\bibnamefont
  {Dietrich}}, \bibinfo {author} {\bibfnamefont {G.}~\bibnamefont {Volpe}},
  \bibinfo {author} {\bibfnamefont {M.~N.}\ \bibnamefont {Sulaiman}}, \bibinfo
  {author} {\bibfnamefont {D.}~\bibnamefont {Renggli}}, \bibinfo {author}
  {\bibfnamefont {I.}~\bibnamefont {Buttinoni}},\ and\ \bibinfo {author}
  {\bibfnamefont {L.}~\bibnamefont {Isa}},\ }\bibfield  {title} {\bibinfo
  {title} {Active atoms and interstitials in two-dimensional colloidal
  crystals},\ }\href {https://doi.org/10.1103/PhysRevLett.120.268004}
  {\bibfield  {journal} {\bibinfo  {journal} {Phys. Rev. Lett.}\ }\textbf
  {\bibinfo {volume} {120}},\ \bibinfo {pages} {268004} (\bibinfo {year}
  {2018})}\BibitemShut {NoStop}%
\bibitem [{\citenamefont {Kümmel}\ \emph {et~al.}(2015)\citenamefont
  {Kümmel}, \citenamefont {Shabestari}, \citenamefont {Lozano}, \citenamefont
  {Volpe},\ and\ \citenamefont {Bechinger}}]{C5SM00827A}%
  \BibitemOpen
  \bibfield  {author} {\bibinfo {author} {\bibfnamefont {F.}~\bibnamefont
  {Kümmel}}, \bibinfo {author} {\bibfnamefont {P.}~\bibnamefont {Shabestari}},
  \bibinfo {author} {\bibfnamefont {C.}~\bibnamefont {Lozano}}, \bibinfo
  {author} {\bibfnamefont {G.}~\bibnamefont {Volpe}},\ and\ \bibinfo {author}
  {\bibfnamefont {C.}~\bibnamefont {Bechinger}},\ }\bibfield  {title} {\bibinfo
  {title} {Formation{,} compression and surface melting of colloidal clusters
  by active particles},\ }\href {https://doi.org/10.1039/C5SM00827A} {\bibfield
   {journal} {\bibinfo  {journal} {Soft Matter}\ }\textbf {\bibinfo {volume}
  {11}},\ \bibinfo {pages} {6187} (\bibinfo {year} {2015})}\BibitemShut
  {NoStop}%
\bibitem [{\citenamefont {Lozano}\ \emph {et~al.}(2019)\citenamefont {Lozano},
  \citenamefont {Gomez-Solano},\ and\ \citenamefont
  {Bechinger}}]{lozano2019active}%
  \BibitemOpen
  \bibfield  {author} {\bibinfo {author} {\bibfnamefont {C.}~\bibnamefont
  {Lozano}}, \bibinfo {author} {\bibfnamefont {J.~R.}\ \bibnamefont
  {Gomez-Solano}},\ and\ \bibinfo {author} {\bibfnamefont {C.}~\bibnamefont
  {Bechinger}},\ }\bibfield  {title} {\bibinfo {title} {Active particles sense
  micromechanical properties of glasses},\ }\href
  {https://doi.org/10.1038/s41563-019-0446-9} {\bibfield  {journal} {\bibinfo
  {journal} {Nature Materials}\ }\textbf {\bibinfo {volume} {18}},\ \bibinfo
  {pages} {1118} (\bibinfo {year} {2019})}\BibitemShut {NoStop}%
\bibitem [{\citenamefont {Omar}\ \emph {et~al.}(2019)\citenamefont {Omar},
  \citenamefont {Wu}, \citenamefont {Wang},\ and\ \citenamefont
  {Brady}}]{Omar2019}%
  \BibitemOpen
  \bibfield  {author} {\bibinfo {author} {\bibfnamefont {A.~K.}\ \bibnamefont
  {Omar}}, \bibinfo {author} {\bibfnamefont {Y.}~\bibnamefont {Wu}}, \bibinfo
  {author} {\bibfnamefont {Z.~G.}\ \bibnamefont {Wang}},\ and\ \bibinfo
  {author} {\bibfnamefont {J.~F.}\ \bibnamefont {Brady}},\ }\bibfield  {title}
  {\bibinfo {title} {{Swimming to Stability: Structural and Dynamical Control
  via Active Doping}},\ }\href
  {http://pubs.acs.org/doi/10.1021/acsnano.8b07421} {\bibfield  {journal}
  {\bibinfo  {journal} {ACS Nano}\ }\textbf {\bibinfo {volume} {13}},\ \bibinfo
  {pages} {560} (\bibinfo {year} {2019})}\BibitemShut {NoStop}%
\bibitem [{\citenamefont {Ma}\ \emph {et~al.}(2017)\citenamefont {Ma},
  \citenamefont {Lei},\ and\ \citenamefont {Ni}}]{C7SM01730H}%
  \BibitemOpen
  \bibfield  {author} {\bibinfo {author} {\bibfnamefont {Z.}~\bibnamefont
  {Ma}}, \bibinfo {author} {\bibfnamefont {Q.-l.}\ \bibnamefont {Lei}},\ and\
  \bibinfo {author} {\bibfnamefont {R.}~\bibnamefont {Ni}},\ }\bibfield
  {title} {\bibinfo {title} {Driving dynamic colloidal assembly using eccentric
  self-propelled colloids},\ }\href {https://doi.org/10.1039/C7SM01730H}
  {\bibfield  {journal} {\bibinfo  {journal} {Soft Matter}\ }\textbf {\bibinfo
  {volume} {13}},\ \bibinfo {pages} {8940} (\bibinfo {year}
  {2017})}\BibitemShut {NoStop}%
\bibitem [{\citenamefont {Ni}\ \emph {et~al.}(2014)\citenamefont {Ni},
  \citenamefont {Cohen~Stuart}, \citenamefont {Dijkstra},\ and\ \citenamefont
  {Bolhuis}}]{C4SM01015A}%
  \BibitemOpen
  \bibfield  {author} {\bibinfo {author} {\bibfnamefont {R.}~\bibnamefont
  {Ni}}, \bibinfo {author} {\bibfnamefont {M.~A.}\ \bibnamefont
  {Cohen~Stuart}}, \bibinfo {author} {\bibfnamefont {M.}~\bibnamefont
  {Dijkstra}},\ and\ \bibinfo {author} {\bibfnamefont {P.~G.}\ \bibnamefont
  {Bolhuis}},\ }\bibfield  {title} {\bibinfo {title} {Crystallizing hard-sphere
  glasses by doping with active particles},\ }\href
  {https://doi.org/10.1039/C4SM01015A} {\bibfield  {journal} {\bibinfo
  {journal} {Soft Matter}\ }\textbf {\bibinfo {volume} {10}},\ \bibinfo {pages}
  {6609} (\bibinfo {year} {2014})}\BibitemShut {NoStop}%
\bibitem [{\citenamefont {Szakasits}\ \emph {et~al.}(2019)\citenamefont
  {Szakasits}, \citenamefont {Saud}, \citenamefont {Mao},\ and\ \citenamefont
  {Solomon}}]{C9SM01496A}%
  \BibitemOpen
  \bibfield  {author} {\bibinfo {author} {\bibfnamefont {M.~E.}\ \bibnamefont
  {Szakasits}}, \bibinfo {author} {\bibfnamefont {K.~T.}\ \bibnamefont {Saud}},
  \bibinfo {author} {\bibfnamefont {X.}~\bibnamefont {Mao}},\ and\ \bibinfo
  {author} {\bibfnamefont {M.~J.}\ \bibnamefont {Solomon}},\ }\bibfield
  {title} {\bibinfo {title} {Rheological implications of embedded active matter
  in colloidal gels},\ }\href {https://doi.org/10.1039/C9SM01496A} {\bibfield
  {journal} {\bibinfo  {journal} {Soft Matter}\ }\textbf {\bibinfo {volume}
  {15}},\ \bibinfo {pages} {8012} (\bibinfo {year} {2019})}\BibitemShut
  {NoStop}%
\bibitem [{\citenamefont {Szakasits}\ \emph {et~al.}(2017)\citenamefont
  {Szakasits}, \citenamefont {Zhang},\ and\ \citenamefont
  {Solomon}}]{PhysRevLett.119.058001}%
  \BibitemOpen
  \bibfield  {author} {\bibinfo {author} {\bibfnamefont {M.~E.}\ \bibnamefont
  {Szakasits}}, \bibinfo {author} {\bibfnamefont {W.}~\bibnamefont {Zhang}},\
  and\ \bibinfo {author} {\bibfnamefont {M.~J.}\ \bibnamefont {Solomon}},\
  }\bibfield  {title} {\bibinfo {title} {Dynamics of fractal cluster gels with
  embedded active colloids},\ }\href
  {https://doi.org/10.1103/PhysRevLett.119.058001} {\bibfield  {journal}
  {\bibinfo  {journal} {Phys. Rev. Lett.}\ }\textbf {\bibinfo {volume} {119}},\
  \bibinfo {pages} {058001} (\bibinfo {year} {2017})}\BibitemShut {NoStop}%
\bibitem [{\citenamefont {van~der Meer}\ \emph
  {et~al.}(2016{\natexlab{a}})\citenamefont {van~der Meer}, \citenamefont
  {Filion},\ and\ \citenamefont {Dijkstra}}]{C6SM00031B}%
  \BibitemOpen
  \bibfield  {author} {\bibinfo {author} {\bibfnamefont {B.}~\bibnamefont
  {van~der Meer}}, \bibinfo {author} {\bibfnamefont {L.}~\bibnamefont
  {Filion}},\ and\ \bibinfo {author} {\bibfnamefont {M.}~\bibnamefont
  {Dijkstra}},\ }\bibfield  {title} {\bibinfo {title} {Fabricating large
  two-dimensional single colloidal crystals by doping with active particles},\
  }\href {https://doi.org/10.1039/C6SM00031B} {\bibfield  {journal} {\bibinfo
  {journal} {Soft Matter}\ }\textbf {\bibinfo {volume} {12}},\ \bibinfo {pages}
  {3406} (\bibinfo {year} {2016}{\natexlab{a}})}\BibitemShut {NoStop}%
\bibitem [{\citenamefont {Ramananarivo}\ \emph {et~al.}(2019)\citenamefont
  {Ramananarivo}, \citenamefont {Ducrot},\ and\ \citenamefont
  {Palacci}}]{Ramananarivo2019}%
  \BibitemOpen
  \bibfield  {author} {\bibinfo {author} {\bibfnamefont {S.}~\bibnamefont
  {Ramananarivo}}, \bibinfo {author} {\bibfnamefont {E.}~\bibnamefont
  {Ducrot}},\ and\ \bibinfo {author} {\bibfnamefont {J.}~\bibnamefont
  {Palacci}},\ }\bibfield  {title} {\bibinfo {title} {{Activity-controlled
  annealing of colloidal monolayers}},\ }\href@noop {} {\bibfield  {journal}
  {\bibinfo  {journal} {Nat. Commun.}\ }\textbf {\bibinfo {volume} {10}},\
  \bibinfo {pages} {1} (\bibinfo {year} {2019})}\BibitemShut {NoStop}%
\bibitem [{\citenamefont {van~der Meer}\ \emph
  {et~al.}(2016{\natexlab{b}})\citenamefont {van~der Meer}, \citenamefont
  {Dijkstra},\ and\ \citenamefont {Filion}}]{C6SM00700G}%
  \BibitemOpen
  \bibfield  {author} {\bibinfo {author} {\bibfnamefont {B.}~\bibnamefont
  {van~der Meer}}, \bibinfo {author} {\bibfnamefont {M.}~\bibnamefont
  {Dijkstra}},\ and\ \bibinfo {author} {\bibfnamefont {L.}~\bibnamefont
  {Filion}},\ }\bibfield  {title} {\bibinfo {title} {Removing grain boundaries
  from three-dimensional colloidal crystals using active dopants},\ }\href
  {https://doi.org/10.1039/C6SM00700G} {\bibfield  {journal} {\bibinfo
  {journal} {Soft Matter}\ }\textbf {\bibinfo {volume} {12}},\ \bibinfo {pages}
  {5630} (\bibinfo {year} {2016}{\natexlab{b}})}\BibitemShut {NoStop}%
\bibitem [{\citenamefont {Stukowski}(2010)}]{Stukowski2010}%
  \BibitemOpen
  \bibfield  {author} {\bibinfo {author} {\bibfnamefont {A.}~\bibnamefont
  {Stukowski}},\ }\bibfield  {title} {\bibinfo {title} {{Visualization and
  analysis of atomistic simulation data with OVITO–the Open Visualization
  Tool}},\ }\href
  {http://stacks.iop.org/0965-0393/18/i=1/a=015012?key=crossref.6895e2c3bb522d1563fb2e2fe9f22789}
  {\bibfield  {journal} {\bibinfo  {journal} {Model. Simul. Mater. Sci. Eng.}\
  }\textbf {\bibinfo {volume} {18}},\ \bibinfo {pages} {015012} (\bibinfo
  {year} {2010})}\BibitemShut {NoStop}%
\bibitem [{Note1()}]{Note1}%
  \BibitemOpen
  \bibinfo {note} {The length of the boundary for a perfect hexagonal crystal
  containing $N$ particles is fairly well described by the empirical function
  $f(N)=-3.309229+3.47188\protect \tmspace +\thinmuskip {.1667em}N^{\protect
  \frac {1}{2}}$.}\BibitemShut {Stop}%
\bibitem [{\citenamefont {Elgeti}\ and\ \citenamefont
  {Gompper}(2013)}]{Elgeti2013}%
  \BibitemOpen
  \bibfield  {author} {\bibinfo {author} {\bibfnamefont {J.}~\bibnamefont
  {Elgeti}}\ and\ \bibinfo {author} {\bibfnamefont {G.}~\bibnamefont
  {Gompper}},\ }\bibfield  {title} {\bibinfo {title} {{Wall accumulation of
  self-propelled spheres}},\ }\href
  {http://stacks.iop.org/0295-5075/101/i=4/a=48003?key=crossref.c1f1a175ed41b22f80ac2d7f2ac7e429}
  {\bibfield  {journal} {\bibinfo  {journal} {EPL}\ }\textbf {\bibinfo {volume}
  {101}},\ \bibinfo {pages} {48003} (\bibinfo {year} {2013})}\BibitemShut
  {NoStop}%
\bibitem [{\citenamefont {Leite}\ \emph {et~al.}(2016)\citenamefont {Leite},
  \citenamefont {Lucena}, \citenamefont {Potiguar},\ and\ \citenamefont
  {Ferreira}}]{leite_depletion_2016}%
  \BibitemOpen
  \bibfield  {author} {\bibinfo {author} {\bibfnamefont {L.~R.}\ \bibnamefont
  {Leite}}, \bibinfo {author} {\bibfnamefont {D.}~\bibnamefont {Lucena}},
  \bibinfo {author} {\bibfnamefont {F.~Q.}\ \bibnamefont {Potiguar}},\ and\
  \bibinfo {author} {\bibfnamefont {W.~P.}\ \bibnamefont {Ferreira}},\
  }\bibfield  {title} {\bibinfo {title} {{Depletion forces on circular and
  elliptical obstacles induced by active matter}},\ }\href
  {https://link.aps.org/doi/10.1103/PhysRevE.94.062602} {\bibfield  {journal}
  {\bibinfo  {journal} {Phys. Rev. E}\ }\textbf {\bibinfo {volume} {94}},\
  \bibinfo {pages} {62602} (\bibinfo {year} {2016})}\BibitemShut {NoStop}%
\bibitem [{\citenamefont {Harder}\ \emph {et~al.}(2014)\citenamefont {Harder},
  \citenamefont {Mallory}, \citenamefont {Tung}, \citenamefont {Valeriani},\
  and\ \citenamefont {Cacciuto}}]{Harder2014}%
  \BibitemOpen
  \bibfield  {author} {\bibinfo {author} {\bibfnamefont {J.}~\bibnamefont
  {Harder}}, \bibinfo {author} {\bibfnamefont {S.~A.}\ \bibnamefont {Mallory}},
  \bibinfo {author} {\bibfnamefont {C.}~\bibnamefont {Tung}}, \bibinfo {author}
  {\bibfnamefont {C.}~\bibnamefont {Valeriani}},\ and\ \bibinfo {author}
  {\bibfnamefont {A.}~\bibnamefont {Cacciuto}},\ }\bibfield  {title} {\bibinfo
  {title} {{The role of particle shape in active depletion}},\ }\href
  {http://aip.scitation.org/doi/full/10.1063/1.4900720
  http://aip.scitation.org/doi/pdf/10.1063/1.4900720
  http://aip.scitation.org/doi/abs/10.1063/1.4900720} {\bibfield  {journal}
  {\bibinfo  {journal} {J. Chem. Phys.}\ }\textbf {\bibinfo {volume} {141}},\
  \bibinfo {pages} {194901} (\bibinfo {year} {2014})}\BibitemShut {NoStop}%
\bibitem [{\citenamefont {Fily}\ \emph {et~al.}(2014)\citenamefont {Fily},
  \citenamefont {Baskaran},\ and\ \citenamefont {Hagan}}]{Fily2014}%
  \BibitemOpen
  \bibfield  {author} {\bibinfo {author} {\bibfnamefont {Y.}~\bibnamefont
  {Fily}}, \bibinfo {author} {\bibfnamefont {A.}~\bibnamefont {Baskaran}},\
  and\ \bibinfo {author} {\bibfnamefont {M.~F.}\ \bibnamefont {Hagan}},\
  }\bibfield  {title} {\bibinfo {title} {{Dynamics of self-propelled particles
  under strong confinement}},\ }\href {http://xlink.rsc.org/?DOI=C4SM00975D
  http://pubs.rsc.org/en/content/articlepdf/2014/sm/c4sm00975d
  http://pubs.rsc.org/en/Content/ArticleLanding/2014/SM/C4SM00975D{\#}!divAbstract}
  {\bibfield  {journal} {\bibinfo  {journal} {Soft Matter}\ }\textbf {\bibinfo
  {volume} {10}},\ \bibinfo {pages} {5609} (\bibinfo {year}
  {2014})}\BibitemShut {NoStop}%
\bibitem [{\citenamefont {Smallenburg}\ and\ \citenamefont
  {L{\"{o}}wen}(2015)}]{Smallenburg2015}%
  \BibitemOpen
  \bibfield  {author} {\bibinfo {author} {\bibfnamefont {F.}~\bibnamefont
  {Smallenburg}}\ and\ \bibinfo {author} {\bibfnamefont {H.}~\bibnamefont
  {L{\"{o}}wen}},\ }\bibfield  {title} {\bibinfo {title} {{Swim pressure on
  walls with curves and corners}},\ }\href
  {https://link.aps.org/doi/10.1103/PhysRevE.92.032304} {\bibfield  {journal}
  {\bibinfo  {journal} {Phys. Rev. E - Stat. Nonlinear, Soft Matter Phys.}\
  }\textbf {\bibinfo {volume} {92}},\ \bibinfo {pages} {32304} (\bibinfo {year}
  {2015})}\BibitemShut {NoStop}%
\bibitem [{\citenamefont {Ray}\ \emph {et~al.}(2014)\citenamefont {Ray},
  \citenamefont {Reichhardt},\ and\ \citenamefont {Reichhardt}}]{Ray2014}%
  \BibitemOpen
  \bibfield  {author} {\bibinfo {author} {\bibfnamefont {D.}~\bibnamefont
  {Ray}}, \bibinfo {author} {\bibfnamefont {C.}~\bibnamefont {Reichhardt}},\
  and\ \bibinfo {author} {\bibfnamefont {C.~J.}\ \bibnamefont {Reichhardt}},\
  }\bibfield  {title} {\bibinfo {title} {{Casimir effect in active matter
  systems}},\ }\href
  {http://arxiv.org/abs/1402.6372{\%}0Ahttp://dx.doi.org/10.1103/PhysRevE.90.013019}
  {\bibfield  {journal} {\bibinfo  {journal} {Phys. Rev. E - Stat. Nonlinear,
  Soft Matter Phys.}\ }\textbf {\bibinfo {volume} {90}},\ \bibinfo {pages}
  {13019} (\bibinfo {year} {2014})}\BibitemShut {NoStop}%
\bibitem [{\citenamefont {Mallory}\ \emph
  {et~al.}(2014{\natexlab{a}})\citenamefont {Mallory}, \citenamefont
  {Valeriani},\ and\ \citenamefont {Cacciuto}}]{Mallory2014a}%
  \BibitemOpen
  \bibfield  {author} {\bibinfo {author} {\bibfnamefont {S.~A.}\ \bibnamefont
  {Mallory}}, \bibinfo {author} {\bibfnamefont {C.}~\bibnamefont {Valeriani}},\
  and\ \bibinfo {author} {\bibfnamefont {A.}~\bibnamefont {Cacciuto}},\
  }\bibfield  {title} {\bibinfo {title} {{Curvature-induced activation of a
  passive tracer in an active bath}},\ }\href
  {https://journals.aps.org/pre/abstract/10.1103/PhysRevE.90.032309} {\bibfield
   {journal} {\bibinfo  {journal} {Phys. Rev. E - Stat. Nonlinear, Soft Matter
  Phys.}\ }\textbf {\bibinfo {volume} {90}},\ \bibinfo {pages} {32309}
  (\bibinfo {year} {2014}{\natexlab{a}})}\BibitemShut {NoStop}%
\bibitem [{\citenamefont {Ni}\ \emph {et~al.}(2015)\citenamefont {Ni},
  \citenamefont {{Cohen Stuart}},\ and\ \citenamefont {Bolhuis}}]{Ni2015}%
  \BibitemOpen
  \bibfield  {author} {\bibinfo {author} {\bibfnamefont {R.}~\bibnamefont
  {Ni}}, \bibinfo {author} {\bibfnamefont {M.~A.}\ \bibnamefont {{Cohen
  Stuart}}},\ and\ \bibinfo {author} {\bibfnamefont {P.~G.}\ \bibnamefont
  {Bolhuis}},\ }\bibfield  {title} {\bibinfo {title} {{Tunable long range
  forces mediated by self-propelled colloidal hard spheres}},\ }\href
  {https://link.aps.org/doi/10.1103/PhysRevLett.114.018302
  https://journals.aps.org/prl/abstract/10.1103/PhysRevLett.114.018302}
  {\bibfield  {journal} {\bibinfo  {journal} {Phys. Rev. Lett.}\ }\textbf
  {\bibinfo {volume} {114}},\ \bibinfo {pages} {18302} (\bibinfo {year}
  {2015})}\BibitemShut {NoStop}%
\bibitem [{\citenamefont {Yan}\ and\ \citenamefont
  {Brady}(2015)}]{yan2015force}%
  \BibitemOpen
  \bibfield  {author} {\bibinfo {author} {\bibfnamefont {W.}~\bibnamefont
  {Yan}}\ and\ \bibinfo {author} {\bibfnamefont {J.~F.}\ \bibnamefont
  {Brady}},\ }\bibfield  {title} {\bibinfo {title} {{The force on a boundary in
  active matter}},\ }\href@noop {} {\bibfield  {journal} {\bibinfo  {journal}
  {J. Fluid Mech.}\ }\textbf {\bibinfo {volume} {785}},\ \bibinfo {pages}
  {R1.1} (\bibinfo {year} {2015})}\BibitemShut {NoStop}%
\bibitem [{\citenamefont {Moran}\ and\ \citenamefont
  {Posner}(2017)}]{Moran2017}%
  \BibitemOpen
  \bibfield  {author} {\bibinfo {author} {\bibfnamefont {J.~L.}\ \bibnamefont
  {Moran}}\ and\ \bibinfo {author} {\bibfnamefont {J.~D.}\ \bibnamefont
  {Posner}},\ }\bibfield  {title} {\bibinfo {title} {Phoretic
  self-propulsion},\ }\href
  {https://doi.org/10.1146/annurev-fluid-122414-034456} {\bibfield  {journal}
  {\bibinfo  {journal} {Annual Review of Fluid Mechanics}\ }\textbf {\bibinfo
  {volume} {49}},\ \bibinfo {pages} {511} (\bibinfo {year} {2017})}\BibitemShut
  {NoStop}%
\bibitem [{\citenamefont {Popescu}\ \emph {et~al.}(2018)\citenamefont
  {Popescu}, \citenamefont {Uspal}, \citenamefont {Domínguez},\ and\
  \citenamefont {Dietrich}}]{Popescu2018}%
  \BibitemOpen
  \bibfield  {author} {\bibinfo {author} {\bibfnamefont {M.~N.}\ \bibnamefont
  {Popescu}}, \bibinfo {author} {\bibfnamefont {W.~E.}\ \bibnamefont {Uspal}},
  \bibinfo {author} {\bibfnamefont {A.}~\bibnamefont {Domínguez}},\ and\
  \bibinfo {author} {\bibfnamefont {S.}~\bibnamefont {Dietrich}},\ }\bibfield
  {title} {\bibinfo {title} {Effective interactions between chemically active
  colloids and interfaces},\ }\href
  {https://doi.org/10.1021/acs.accounts.8b00237} {\bibfield  {journal}
  {\bibinfo  {journal} {Accounts of Chemical Research}\ }\textbf {\bibinfo
  {volume} {51}},\ \bibinfo {pages} {2991} (\bibinfo {year}
  {2018})}\BibitemShut {NoStop}%
\bibitem [{\citenamefont {Vutukuri}\ \emph {et~al.}(2020)\citenamefont
  {Vutukuri}, \citenamefont {Lisicki}, \citenamefont {Lauga},\ and\
  \citenamefont {Vermant}}]{Vutukuri2020}%
  \BibitemOpen
  \bibfield  {author} {\bibinfo {author} {\bibfnamefont {H.~R.}\ \bibnamefont
  {Vutukuri}}, \bibinfo {author} {\bibfnamefont {M.}~\bibnamefont {Lisicki}},
  \bibinfo {author} {\bibfnamefont {E.}~\bibnamefont {Lauga}},\ and\ \bibinfo
  {author} {\bibfnamefont {J.}~\bibnamefont {Vermant}},\ }\bibfield  {title}
  {\bibinfo {title} {Light-switchable propulsion of active particles with
  reversible interactions},\ }\href
  {https://doi.org/10.1038/s41467-020-15764-1} {\bibfield  {journal} {\bibinfo
  {journal} {Nature Communications}\ }\textbf {\bibinfo {volume} {11}},\
  \bibinfo {pages} {1} (\bibinfo {year} {2020})}\BibitemShut {NoStop}%
\bibitem [{\citenamefont {Takatori}\ \emph {et~al.}(2014)\citenamefont
  {Takatori}, \citenamefont {Yan},\ and\ \citenamefont {Brady}}]{Takatori2014}%
  \BibitemOpen
  \bibfield  {author} {\bibinfo {author} {\bibfnamefont {S.~C.}\ \bibnamefont
  {Takatori}}, \bibinfo {author} {\bibfnamefont {W.}~\bibnamefont {Yan}},\ and\
  \bibinfo {author} {\bibfnamefont {J.~F.}\ \bibnamefont {Brady}},\ }\bibfield
  {title} {\bibinfo {title} {{Swim pressure: Stress generation in active
  matter}},\ }\href {https://link.aps.org/doi/10.1103/PhysRevLett.113.028103
  https://journals.aps.org/prl/abstract/10.1103/PhysRevLett.113.028103}
  {\bibfield  {journal} {\bibinfo  {journal} {Phys. Rev. Lett.}\ }\textbf
  {\bibinfo {volume} {113}},\ \bibinfo {pages} {28103} (\bibinfo {year}
  {2014})}\BibitemShut {NoStop}%
\bibitem [{\citenamefont {Mallory}\ \emph
  {et~al.}(2014{\natexlab{b}})\citenamefont {Mallory}, \citenamefont {\ifmmode
  \check{S}\else \v{S}\fi{}ari\ifmmode~\acute{c}\else \'{c}\fi{}},
  \citenamefont {Valeriani},\ and\ \citenamefont {Cacciuto}}]{Mallory2014b}%
  \BibitemOpen
  \bibfield  {author} {\bibinfo {author} {\bibfnamefont {S.~A.}\ \bibnamefont
  {Mallory}}, \bibinfo {author} {\bibfnamefont {A.}~\bibnamefont {\ifmmode
  \check{S}\else \v{S}\fi{}ari\ifmmode~\acute{c}\else \'{c}\fi{}}}, \bibinfo
  {author} {\bibfnamefont {C.}~\bibnamefont {Valeriani}},\ and\ \bibinfo
  {author} {\bibfnamefont {A.}~\bibnamefont {Cacciuto}},\ }\bibfield  {title}
  {\bibinfo {title} {Anomalous thermomechanical properties of a self-propelled
  colloidal fluid},\ }\href {https://doi.org/10.1103/PhysRevE.89.052303}
  {\bibfield  {journal} {\bibinfo  {journal} {Phys. Rev. E}\ }\textbf {\bibinfo
  {volume} {89}},\ \bibinfo {pages} {052303} (\bibinfo {year}
  {2014}{\natexlab{b}})}\BibitemShut {NoStop}%
\bibitem [{\citenamefont {Fily}\ and\ \citenamefont
  {Marchetti}(2012)}]{Fily2012}%
  \BibitemOpen
  \bibfield  {author} {\bibinfo {author} {\bibfnamefont {Y.}~\bibnamefont
  {Fily}}\ and\ \bibinfo {author} {\bibfnamefont {M.~C.}\ \bibnamefont
  {Marchetti}},\ }\bibfield  {title} {\bibinfo {title} {{Athermal phase
  separation of self-propelled particles with no alignment}},\ }\href@noop {}
  {\bibfield  {journal} {\bibinfo  {journal} {Phys. Rev. Lett.}\ }\textbf
  {\bibinfo {volume} {108}},\ \bibinfo {pages} {235702} (\bibinfo {year}
  {2012})}\BibitemShut {NoStop}%
\bibitem [{\citenamefont {{Zaeifi Yamchi}}\ and\ \citenamefont
  {Naji}(2017)}]{ZaeifiYamchi2017}%
  \BibitemOpen
  \bibfield  {author} {\bibinfo {author} {\bibfnamefont {M.}~\bibnamefont
  {{Zaeifi Yamchi}}}\ and\ \bibinfo {author} {\bibfnamefont {A.}~\bibnamefont
  {Naji}},\ }\bibfield  {title} {\bibinfo {title} {{Effective interactions
  between inclusions in an active bath}},\ }\href@noop {} {\bibfield  {journal}
  {\bibinfo  {journal} {J. Chem. Phys.}\ }\textbf {\bibinfo {volume} {147}},\
  \bibinfo {pages} {194901} (\bibinfo {year} {2017})}\BibitemShut {NoStop}%
\bibitem [{\citenamefont {Mognetti}\ \emph {et~al.}(2013)\citenamefont
  {Mognetti}, \citenamefont {{\v{S}}ari{\'{c}}}, \citenamefont
  {Angioletti-Uberti}, \citenamefont {Cacciuto}, \citenamefont {Valeriani},\
  and\ \citenamefont {Frenkel}}]{mognetti_living_2013}%
  \BibitemOpen
  \bibfield  {author} {\bibinfo {author} {\bibfnamefont {B.~M.}\ \bibnamefont
  {Mognetti}}, \bibinfo {author} {\bibfnamefont {A.}~\bibnamefont
  {{\v{S}}ari{\'{c}}}}, \bibinfo {author} {\bibfnamefont {S.}~\bibnamefont
  {Angioletti-Uberti}}, \bibinfo {author} {\bibfnamefont {A.}~\bibnamefont
  {Cacciuto}}, \bibinfo {author} {\bibfnamefont {C.}~\bibnamefont
  {Valeriani}},\ and\ \bibinfo {author} {\bibfnamefont {D.}~\bibnamefont
  {Frenkel}},\ }\bibfield  {title} {\bibinfo {title} {{Living clusters and
  crystals from low-density suspensions of active colloids}},\ }\href
  {https://link.aps.org/doi/10.1103/PhysRevLett.111.245702} {\bibfield
  {journal} {\bibinfo  {journal} {Phys. Rev. Lett.}\ }\textbf {\bibinfo
  {volume} {111}},\ \bibinfo {pages} {245702} (\bibinfo {year}
  {2013})}\BibitemShut {NoStop}%
\end{thebibliography}

%apsrev4-2.bst 2019-01-14 (MD) hand-edited version of apsrev4-1.bst
%Control: key (0)
%Control: author (8) initials jnrlst
%Control: editor formatted (1) identically to author
%Control: production of article title (0) allowed
%Control: page (0) single
%Control: year (1) truncated
%Control: production of eprint (0) enabled
%

\end{document}